\documentclass[aps,pre,reprint,amsmath,amssymb,graphicx,longbibliography,superscriptaddress]{revtex4-1}

\usepackage{bm}

\usepackage{relsize}
\usepackage{simplewick}
\usepackage{graphicx}

\usepackage{indentfirst}
\usepackage{psfrag}
\usepackage{epsfig}
\usepackage{color}

\newcommand{\rr}{{\bm{\rho}}}

\def\MT{\widehat{\mathbb{T}}}
\def\be{\begin{eqnarray}}
\def\ee{\end{eqnarray}}
\begin{document}

\title{Central limit theorem for the wave transport in disordered waveguides: a perturbative approach}

\author{M. Y\'epez}
\affiliation{Departamento de F\'isica, Universidad Aut\'onoma Metropolitana-Iztapalapa, A. P. 55-534, 09340 M\'exico Distrito Federal, M\'exico.}

\author{J.J. S\'aenz}
\affiliation{Donostia International Physics Center (DIPC), Paseo Manuel Lardizabal 4, 20018 Donostia-San Sebastian, Spain.}
\affiliation{IKERBASQUE, Basque Foundation for Science, 48013 Bilbao, Spain.}

\date{\today}

\begin{abstract}

The statistical scattering properties of wave transport in disordered waveguides are derived perturbatively within the transition matrix formalism. The limiting macroscopic statistic of the wave transport, emerges as a consequence of a generalized central-limit-theorem: the expectation values of macroscopic observables depend only on the first and second moments of the reflection matrix of individual scatterers. This theoretical approach does not consider any statistical assumption on the scattering properties of the individual scatterers and it is fully consistent with the optical theorem. The results are found in the so called dense-weak-scattering limit and in the ballistic regime.

\end{abstract}



\maketitle

\section{Introduction}

Scaling and universality \cite{abrahams1979scaling,*anderson1980new} are fundamental  concepts in the description of wave transport through random and complex media \cite{Sheng:1990,*van2003wave,Mello:2010}. The basic physical assumption  underlying these concepts is that the transport properties of systems probed at length scales much larger than the {\em transport mean free path}, $\ell$, should be insensitive to the microscopic details needed to specify completely the system. Remarkable  phenomena such as coherent enhanced backscattering \cite{kuga1984retroreflectance,*van1985observation,*wolf1985weak}, universal conductance fluctuations \cite{lee1985universal,*altshuler1985fluctuations,*imry1986active,*mello1988macroscopic2,*scheffold1998universal} and the underlying correlations between transport coefficients \cite{feng1988correlations,*mello1988macroscopic,*feng1991mesoscopic,*sebbah2002spatial,garcia2002finite,*cwilich2006spatial,*froufe2007fluctuations}, or the shape of the conductance distributions in the diffusive and localisation regimes \cite{PhysRevLett_87_116603,PhysRevLett_89_246403} were found to be independent of the  microscopic details.

In general, the statistical properties of wave transport \cite{PhysRevB.37.5860,AnnaPhys.181.290,PismaZh.36.259,*JETPLett.36.318,PhysRevB.46.15963,WRCM.15.229,PhysRevLett_87_116603,PhysRevLett_89_246403,PhysRevE.75.031113,PhysRevB.83.245328,EPL.108.17006,*YepezSaenz:arXiv:1407.5617} present significant regularities in the sense that the {\em limiting macroscopic statistic} of macroscopic observables involves a rather small number of physical parameters. This signals the existence of a generalised {\em central limit theorem} (CLT), in analogy with the  familiar CLT in statistics: once the scaling parameters are specified, the limiting statistics is {\em universal}.

Early work had already shown that a certain class of models, in the {\em dense weak scattering limit} (DWSL), displays a universal limiting distribution specified by the mean free path (MFP) as the only scaling parameter, confirming the validity of the scaling hypothesis. This result was obtained by using the DMPK  multichannel scattering approach (after Dorokhov \cite{PismaZh.36.259,*JETPLett.36.318} and Mello, Pereyra and Kumar \cite{AnnaPhys.181.290}). The same results were obtained from the non-linear sigma model \cite{JETPLett.58.444,*IntJMP.8.3795}, which was shown to be equivalent to the DMPK \cite{PhysRevB.53.1490}. In spite of the successes of these approaches, the DMPK theoretical description of the scattering processes is not fully rigorous since it does not fulfil the Optical Theorem (OT): A key assumption of these scaling approaches is the hypothesis of  isotropic distribution of phases or {\em isotropy hypothesis} (i.e. the phases of the scattering and transfer matrices are statistically independent and uniformly distributed \cite{JMP.27.2876,PhysRevB.44.3559}), which imposes a null value for the scattering amplitude in the forward direction, while the average of the total scattered flux is not zero, in disagreement with the OT. Our main goal here is to  demonstrate the existence of a limiting macroscopic statistic, consistent with the OT, emerging as consequence of a generalized Central Limit Theorem in the ballistic regime.

Most previous studies \cite{PhysRevB.37.5860,AnnaPhys.181.290,PismaZh.36.259,*JETPLett.36.318} were mainly focused on the diffusive $\ell \ll L \ll \xi $ and localised $\xi \ll L$ regimes, where $L$ denotes the length of the disordered system, while  $\xi$ denote the localisation length. When the scattering is isotropic (e.g. scattering units smaller than the wavelength and uniformly distributed through the bulk of the system), the excellent agreement of the DMPK predictions concerning the full conductance distributions in the crossover region between diffusive  and localised regimes \cite{PhysRevLett_89_246403}, suggests that, once the system is deep inside the diffusive regime, i.e. $L \gg \ell$, the isotropy hypothesis is a very good approximation although, from a fundamental point of view, not fully satisfactory. Even in the  crossover from the ballistic to diffusive  regimes, $L \lesssim \ell$,  the DMPK scaling approach is able to capture the transition from negative to positive intensity-intensity correlations \cite{garcia2002finite,*cwilich2006spatial,*froufe2007fluctuations} and the conductance fluctuations at subdiffusion scales \cite{garcia2002finite,*cwilich2006spatial,
*froufe2007fluctuations,Marinyuk2015mesoscopic}.

The isotropy hypothesis can not be apply if one is interested in situations when the scattering is not isotropic, as in samples with rough surfaces and no bulk disorder \cite{WRCM.15.229}. A clear example was given in \cite{PhysRevLett_89_246403} where the conductance distribution was studied in the crossover region between diffusive  and localised regimes. For waveguides with bulk disorder the description is excellent \cite{PhysRevLett_89_246403}, whereas it fails to describe the results of waveguides with surface disorder \cite{PhysRevLett_87_116603}. This has motivated several attempts to extend the CLT beyond the simplified hypothesis of isotropy. In reference \cite{PhysRevB.46.15963} a limiting distribution wider than that of Ref. \cite{AnnaPhys.181.290} was found by Mello and Tomsovic (MT), in which the ``isotropy'' of phases was not implemented to a large extent. The central result was also found in the DWSL, where an evolution equation for the expectation value of macroscopic observables was derived. The MT description, depends only on the microscopic details through the channel-channel $\ell_{aa_{0}}$ and the scattering $\ell_{a_{0}}$ mean free paths for the various scattering processes that may occur. As an extension of these results, a CLT was found \cite{PhysRevE.75.031113} using a Fokker-Planck description of the evolution of the expectation value of different physical quantities as a function of the length of the system. An important improvement of Ref. \cite{PhysRevE.75.031113} compared with the MT result was that the energy of the incident particle was fully taken into account. The introduction of the different mean free paths $\ell_{aa_{0}}$ and $\ell_{a_{0}}$ capture the physics when the difference in behavior of the various modes becomes relevant in the wave transport, as it is the case of the wave transport through waveguides with surface disorder \cite{WRCM.15.229} where the DMPK approach does not give a suitable description. However, none of these previous works provide a CLT fully consistent with the conditions imposed by the Optical Theorem. As we will show here, a consistent approach to obtain the limiting macroscopic statistics requires a proper description of the role of evanescent modes; for instance, in Ref. \cite{EPL.108.17006} it is shown that the channel dependent effective wave number $k_{\mathrm{eff}}$ of disordered waveguides, is quite sensitive to the influence of the evanescent modes.
 
The motivation of the present work is to study the statistical scattering properties of disordered systems in the ballistic regime where the scattering anisotropies and the Optical Theorem plays a transcendental role. The aim is to demonstrate the existence of a limiting macroscopic statistics, which emerges from a more general central limit theorem than those ones found in previous works. Since in the ballistic regime the lowest order processes in multiple scattering dominate the response of the system, we use perturbation theory to study the statistical scattering properties of the system within  the {\em transition matrix} formalism \cite{AnndAstPhy.30.565,Roman:1965,*Newton:1982,*Messiah:1999}. We show the existence of a generalized Central Limit Theorem, in the sense the expectation values depend on the microscopic details of the disorder through the first and second moments of the extended reflection matrix elements of the individual scatterers. The CLT found in the present work, present important fundamental differences with respect to the CLT's found in previous works: i) The energy of the incident particle is fully taken into account, ii) This CLT allows us to identify the order in multiple scattering of the wave transport, where open and closed channels transitions are included through the statistical moments  iii) The derivation of the present CLT, does not contain any assumption on the statistical scattering properties of the  individual scattering units and iv) the results are fully consistent with the Optical Theorem.

The paper is organized as follows. In Sec. \ref{summery_MyPHDCap2} we describe the theoretical model of the disordered system and summarise its statistical scattering properties. The scattering problem is defined in the transition matrix formalism and the Flux Conservation property and Optical Theorem are also introduced. In Sec. \ref{LimitingStatist}, the statistical properties of the wave transport of the disordered system are studied perturbatively by using the transition matrix method. The resulting series expansion relates the statistical average of the macroscopic observables of interest, with the microscopic statistic of the disordered system, what allows us to identify the role of the multiple scattering processes in the macroscopic observables of interest. In Sec. \ref{LimitingStatist}, we also obtain, in the {\em dense-weak-scattering limit} and in the ballistic regime, the generalized central-limit theorem, that gives rise to the limiting macroscopic statistics of the wave transport, where only the first and second statistical moments of the extended reflection matrix of the single scatterers are relevant, while higher order moments play no role. In Sec. \ref{Gen_CLT} the scaling parameters of previous descriptions are recovered in the so called {\em short-wavelength approximation} and when the evanescent transport is not relevant. We finally conclude in Sec. \ref{Conclusions}.

\section{\label{summery_MyPHDCap2}The disordered system and its scattering properties}

Here we introduce our model system and summarise  some key concepts of the wave scattering in a {\it quasi-one dimensional} (Q1D) disordered systems, where the propagation is constricted to one direction. We consider the wave equation
\be
\nabla ^{2} \Psi(x,y) + k^{2} \Psi = U(x,y) \Psi(x,y),
\label{WE}
\ee
inside a waveguide with impenetrable walls and uniform cross section $W$, so boundary conditions impose $\Psi \left(x,0 \right)=\Psi\left(x, W\right)=0$. A bulk disordered region of  length $L$, divides the waveguide into two, otherwise clean ($U=0$ ), left ($x<0$) and right ($x>L$) leads: see Fig. \ref{Schematic}. The lateral confinement defines a set
$\{ \chi_b(y); b=1,2,\cdots \}$ of orthonormal transversal eigenfunctions
\begin{equation}
\chi_b(y) = \sqrt{\frac{2}{W}} \sin \frac{\pi b y}{W}.
\label{Tran_Mods} 
\end{equation}
The asymptotic wave function deep inside the leads can be described as a linear combination of propagating or traveling modes 
\begin{equation}
\Phi_{\sigma b}(\rr) = \frac{e^{i\sigma k_b x}}{\sqrt{k_b}} \chi_b(y) = \varphi_{\sigma}(k_b;x) \chi_b(y),
\end{equation}
where $k_{b}=\sqrt{k^{2}-(\pi b/W)^{2}}$ is the real longitudinal wave number and $\sigma = \pm$ specifies the propagation direction in the open channel $b$ ($\sigma = +$ for traveling from left to right). The waveguide supports precisely $N$ traveling modes or open channels ($1\leq b\leq N $), when $N < kW/ \pi < N + 1$, $k=2\pi/\lambda$ being the wave number in the clean regions of the waveguide; modes with $b > N$, represent evanescent modes with imaginary longitudinal wave number $k_{b}\rightarrow i\kappa_{b}$, being $\kappa_{b}=\sqrt{(\pi b/W)^{2}-k^{2}}$. 

\subsection{\label{Generalities}Scattering Theory in a 2D waveguide}

Consider an incoming wave in the open channel $a_0$, $\Phi_{\sigma a_0}(\rr)$. The wave equation \eqref{WE}, can be written as a Lippmann-Schwinger integral equation
\begin{align}
\Psi_{\sigma a_0} (\rr) = \Phi_{\sigma a_0}(\rr) - \int G_0(\rr;\rr') U(\rr') \Psi_{\sigma a_0} (\rr') d^2\rr' ,
\label{LS}
\end{align}
being $G_0(\rr,\rr')$ the Green function of the clean waveguide. The expansion of $\Psi_{\sigma a_{0}}\left(\rr \right)$ and $G_0(\rr,\rr')$ in the basis of transversal modes,
\be
\Psi_{\sigma a_{0}}\left(\rr \right) &=& \sum_{b=1}^{\infty} \left[ \psi_{\sigma a_{0}}\left(x\right) \right]_{b} \chi_{b}\left( y \right),
\label{MoreGenSolShrEq} \\
G_0(\rr;\rr') &=& \sum_{b=1}^\infty \frac{i}{2k_b}e^{ik_b|x-x' |} \chi_b(y) \chi_{b}(y'),
\ee
allow us to transform Eq. \eqref{LS} into a set of coupled equations for the longitudinal components \cite{Mello:2010, Roman:1965,*Newton:1982,*Messiah:1999,mosk1996theory}:
\begin{align}
\left[ \psi_{\sigma a_{0}}\left(x\right) \right]_{b}& =\varphi _{\sigma}\left( k_{b};x\right) \delta_{ba_{0}}
\label{MY_PHDT_2_28ab}
\\
- \int_0^L &\sum_{b_{1}=1}^{\infty} \left[{\bm{G}}_0 (x,x') \right]_{bb} \left[ {\bf{u}}(x')\right] _{bb_{1}}
\left[ \psi _{\sigma a_{0}}\left(x'\right) \right]_{b_{1}} dx' , \nonumber
\end{align}
where we have defined the matrix elements
\be
[{\bf{u}}(x)]_{bb'} &=& \int_0^W \chi_b(y) U(x,y) \chi_{b'}(y)  dy, \label{umatrix} \\
\left[{\bm{G}}_0 (x,x') \right]_{bb'} &=& \delta_{bb'} \frac{i}{2k_b}e^{ik_b|x-x'|}. \label{GFb}
\ee

The set of coupled equations of Eq. \eqref{MY_PHDT_2_28ab} can be rewritten in the following compact vectorial form
\be
\bm{\psi}_{\sigma}\left(x\right) = \bm{\varphi} _{\sigma}\left(x\right) 
-\int_0^L \bm{G}_0\left(x,x' \right) {\bf{u}}(x')
\bm{\psi} _{\sigma }\left(x'\right)  dx',
\label{CompactLimppman}
\ee
or equivalently
\begin{equation}
\bm{\psi}_{\sigma}\left(x\right) = \bm{\varphi} _{\sigma}\left(x\right) 
-
\iint_{0}^{L}
\bm{G}_0\left(x,x' \right) \widehat{ \mathbb{T} } \left( x^{\prime },x^{\prime \prime}\right)
\bm{\varphi} _{\sigma}\left(x''\right)  dx'dx'',
\label{TMT}
\end{equation}
where $\widehat{ \mathbb{T} }\left( x^{\prime },x^{\prime \prime}\right)$ denotes the  transition matrix operator.

\subsubsection{Scattering and Transition matrices in a waveguide}

The wave transport properties are described by means of the well known $2N \times 2N $ {\em open channels} or {\em reduced scattering matrix}

\begin{equation}
\bm{S}=
\left(
\begin{array}{cc}
\bm{r} & \bm{t}^{\prime}
\\
\bm{t} & \bm{r}^{\prime}
\end{array}
\right),
\label{ScatMat_Relat}
\end{equation}
\noindent which relates open channel outgoing- and incoming-wave amplitudes in the asymptotic region. The reflection, $r_{aa_0}$, and transmission, $t_{aa_0}$, elements give the asymptotic outgoing wave amplitude in channel $a$ for an incoming ($\sigma = +$) mode in the open channel $a_0$,
\begin{subequations}
\begin{eqnarray}
\left[\psi_{+a_{0}}\left(x\right)\right]_{a} &\overset{x \rightarrow -\infty}{\sim}& \varphi_{+}\left(k_{a};x\right)\delta_{aa_{0}} + r_{aa_{0}}\varphi_{-}\left(k_{a};x\right),
\nonumber
\\
\label{AsympLongComp1}
\\
\left[\psi_{+a_{0}}\left(x\right)\right]_{a}&\overset{x \rightarrow +\infty}{\sim}&  t_{aa_{0}}\varphi_{+}\left(k_{a};x\right);
\label{AsympLongComp2}
\end{eqnarray}
\label{AsympLongComp1_2}
\end{subequations}

\noindent equivalent expressions for the elements of $\bm{r}'$ and $\bm{t}'$ are obtained from waves incoming from right to left.
 
The reflection and transmission amplitudes are related to the elements of the {\em transition matrix} $\bf{T}$,
\begin{eqnarray}
\mathtt{T} _{aa_{0}} ^ {\sigma^{\prime} \sigma} \equiv  \iint_{0}^{L} \varphi_{\sigma^{\prime}} ^{\ast} \left(k_{a};x^{\prime}\right) \MT_{aa_{0}} \left( x^{\prime }, x^{\prime \prime} \right) \varphi_{\sigma} \left(k_{a_{0}};x^{\prime \prime}\right) dx^{\prime \prime} dx^{\prime },
\nonumber \\
\label{ReDef_Back_For_scatt_Ampli}
\end{eqnarray}
 as
\begin{subequations}
\begin{eqnarray}
r_{aa_{0}}&=&-\frac{i}{2} \mathtt{T} _{aa_{0}} ^{ - + },  \quad t_{aa_{0}}=\delta_{aa_{0}}-\frac{i}{2} \mathtt{T} _{aa_{0}} ^{ + + }, \\
  r'_{aa_{0}}&=& -\frac{i}{2}\mathtt{T} _{aa_{0}} ^{ + - },   \quad t'_{aa_{0}} =\delta_{aa_{0}} -\frac{i}{2}\mathtt{T} _{aa_{0}} ^{ - - }.
\end{eqnarray}
\label{ScattAmplandTMatrix}
\end{subequations}

It is important to emphasise that  {\em flux conservation} (FC) lead to the following constrictions
\begin{subequations}
\begin{eqnarray}
T_{a_{0}}+R_{a_{0}}&=&1,
\label{FluxConsRelation}
\\
\;\;
T_{a_{0}}=\sum_{a=1}^{N}T_{aa_{0}},
&\;\;&
R_{a_{0}}=\sum_{a=1}^{N}R_{aa_{0}},
\end{eqnarray}
where $T_{aa_{0}}=\vert t_{aa_{0}}\vert^{2}$ and $R_{aa_{0}}=\vert r_{aa_{0}}\vert^{2}$ denote, respectively, the individual channel-channel transmission and reflection coefficients, while $T_{a_{0}}$ and $R_{a_{0}}$ represent the total transmission and reflection coefficients when the incidence is given in the channel $a_{0}$. In terms of the transition matrix elements, flux conservation implies
 \begin{equation}
 \sum_{a=1}^{N}\left( \vert \mathtt{T}_{aa_{0}}^{- +}
\vert^{2}
+ 
\vert \mathtt{T}_{aa_{0}}^{+ +}
\vert^{2} \right) =
-4 \
\mathrm{Im} \left(\mathtt{T} _{a_{0}a_{0}} ^{+ + }\right),
\label{GeneralOpticalTheorem}
\end{equation}
\label{Both_FC_and_OPTh}
\end{subequations}

\noindent which is also known as the {\em Optical Theorem} (OT). Any possible microscopic realization of the disorder must be consistent with the FC property and the OT given in Eq. \eqref{Both_FC_and_OPTh}.

\subsection{The disordered region}

\begin{figure}[t]
\begin{center}
\includegraphics[width=\columnwidth]{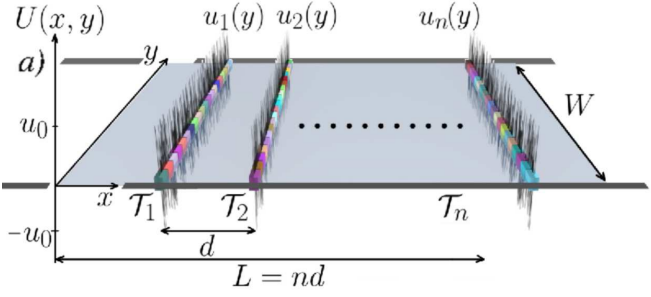}
\caption{\small{ Schematic representation of the Building Block as a sequence of thin potential slices.}}
\label{Schematic}
\end{center}
\end{figure}
The disordered region, which hereafter shall be called the {\em Building Block} (BB), is represented by the random potential $U\left( x,y\right)$ (in units of $k^{2}$) shown schematically in Fig. \ref{Schematic}: The potential is a sequence of $n$ ($n \gg 1$) statistically independent and identically distributed thin scattering units of thickness $\delta$ (with $k \delta  \ll 1$) and separated one each other by a fixed distance $d= f^{-1} \delta$, $f\le 1$ being the filling fraction; therefore, the potential model of the BB is given by
\be
U(x,y) &=& \sum_{r=1}^n U_{r}\left(x,y\right) \nonumber \\
U_{r}\left(x,y\right) 
&=&u_{r}\left(y\right)\Theta\left( \frac{\delta}{2}-\vert x-x_{r} \vert \right), \quad x_r=rd,
\label{MyPHDTEq6.13}
\ee
where $u_r(y)$ is an arbitrary potential profile, whose matrix elements are given by  $({\bf{u}}_r)_{bb'} \equiv \left[ {\bf{u}}(x_r)\right] _{bb'}$ [see Eq. \eqref{umatrix}].

The statistical properties of the potential are contained in the moments of the potential matrix elements. We indicate the $q$th moments as
\be
\left\langle ({\bf{u}}_r)_{b_1b_1^{\prime}} ({\bf{u}}_r)_{b_2b_2^{\prime}}  \cdots ({\bf{u}}_r)_{b_{q}b_{q}^{\prime}}        \right\rangle
= \mu^{({\bf{u}})}_{b_1b_1' \cdots b_q b_q'}.
\ee
We assume, for simplicity, that all the statistical odd-moments of the potential matrix elements vanish
\be
\mu^{({\bf{u}}_r)}_{b_1b_1' \cdots b_{2p-1} b_{2p-1}'} = 0, \quad p=1,2,\cdots .
\label{oddmomentsu}
\ee
Concerning the even moments, we shall find it convenient to make the change of variables 
\be
({\bf{u}}_r)_{bb'} = \frac{1}{\sqrt{\delta}} (\tilde{\bf{u}}_r)_{bb'}, \quad \text{e.g.} \; u_r(y) =  \frac{1}{\sqrt{\delta}} \tilde{u}_r(y),
\ee
and consider $\tilde{u}_r(y) $, the distribution of $(\tilde{\bf{u}}_r)_{bb'} $ and the corresponding moments, $ \mu^{(\tilde{{\bf{u}}}_r)}$, independent of the slice thickness $\delta$. Then, in this statistical model, the even moments scale with $\delta$ as  
\be
\mu^{({\bf{u}}_r)}_{b_1b_1' \cdots b_{2p} b_{2p}'} =  \frac{1}{\delta^p}  \  \mu^{(\tilde{\bf{u}}_r)}_{b_1b_1' \cdots b_{2p} b_{2p}'}  , \quad p=1,2,\cdots .
\label{evenmomentsu}
\ee

\section{\label{LimitingStatist}Generalized central limit theorem}

\subsection{\label{Moments_of_TransitionMatrix}Perturbative approach}

In this section we study perturbatively the statistical scattering properties of the BB in the ballistic regime, where the length $L$ of the BB is much smaller than the transport or elastic mean free path $\ell$ (see Sec. \ref{Gen_CLT}); in this regime, the lowest order contributions in multiple scattering dominate the wave transport through the BB. The aim is to obtain the macroscopic statistics of the BB in terms of its microscopic statistics.

In our perturbative analysis, we first   consider that, in the limit $\delta \ll \lambda$, the wavefunction $\Psi$ is  constant  along the wave propagation direction $x$ inside each potential slice [see Appendix \ref{Tmss}]. The set of 
coupled Lippmann-Schwinger equations, Eq. \eqref{MY_PHDT_2_28ab}, can be formally solved using the standard perturbative technique of Ref. \cite{AnndAstPhy.30.565}.
The perturbative analysis discussed below is focused on the statistical average or expectation values of the complex transition matrix elements $\left\langle \mathtt{T}^{\pm +}_{aa_{0}} \right\rangle_{L}$, while the relevant results for the (real) second moments  $\left\langle \vert \mathtt{T}^{\pm +}_{aa_{0}}  \vert^{2} \right \rangle_{L}$ are only mentioned [see Eq. \eqref{ScattAmplandTMatrix}]. Once the perturbative series expansion for the transition matrix is found [see Appendix \ref{FinalGenScattTheory}], the different orders in multiple scattering are reorganized through the identification of the different orders of multiple scattering between a fixed number of scattering units; this represents the key point of the present perturbative approach.

For a given length $L$ of the disordered region, $\left\langle \mathtt{T}^{\pm +}_{aa_{0}} \right\rangle_{L}$ can be written  as a series expansion in powers of the $\tilde{\bm{r}}_r$-matrices of the single slices 
\begin{widetext}
\begin{subequations}
\be
-\frac{i}{2} \left\langle  \mathtt{T} _{aa_{0}} ^ {\sigma^{\prime} \sigma} \right\rangle_L &&  =
\sum_{r_1}^n \Biggl\langle  
\ _{\sigma'} \! - \! \! \! \!  \overset{\tilde{\bm{r}}_{r_1} }{\mathlarger{\bullet}}  \! \! \! \!- _{\sigma}
\Biggr\rangle_{a a_0} \\
&& + 
\sum_{r_1 \ne r_2}^n \Biggl\langle \ _{\sigma'} \! - \! \! \! \!  \overset{\tilde{\bm{r}}_{r_1} \tilde{\bm{r}}_{r_2} }  
{\mathlarger{\bullet} \! \! \! - \! \! \! - \! \!\mathlarger{\bullet} } \! \! \! \!- _{\sigma}  
+ 
\ _{\sigma'} \! - \! \! \! \!  \overset{\tilde{\bm{r}}_{r_1} \tilde{\bm{r}}_{r_2} \tilde{\bm{r}}_{r_1}}
 {
 \bcontraction{}{ \mathlarger{\bullet} }{ \! \! \!- \! \! \!- \! \!\mathlarger{\bullet}  \! \! \!- \! \! \!- \! \!}{\mathlarger{\bullet}}
 \mathlarger{\bullet} \! \! \!- \! \! \!- \! \!\mathlarger{\bullet}  \! \! \!- \! \! \!- \! \!\mathlarger{\bullet}
 } \! \! \! \!- _{\sigma}
 +  
\ _{\sigma'} \! - \! \! \! \! \overset{\tilde{\bm{r}}_{r_1} \tilde{\bm{r}}_{r_2} \tilde{\bm{r}}_{r_1} \tilde{\bm{r}}_{r_2}}
 {
 \bcontraction[2ex]{}{ \mathlarger{\bullet} }{\! \! \!- \! \! \!- \! \!\mathlarger{\bullet}  \! \! \!- \! \! \!- \! \!}{\mathlarger{\bullet}}
 \bcontraction{ \mathlarger{\bullet} \! \! \!- \! \! \!- \! \!}{\mathlarger{\bullet}} { \! \! \!- \! \! \!- \! \!\mathlarger{\bullet}  \! \! \!- \! \! \!- \! \!}{\mathlarger{\bullet}}
\mathlarger{\bullet} \! \! \!- \! \! \!- \! \!\mathlarger{\bullet}  \! \! \!- \! \! \!- \! \!\mathlarger{\bullet} \! \! \!- \! \! \!- \! \!\mathlarger{\bullet}
 }\! \! \! \!- _{\sigma}
 +
\ _{\sigma'} \! - \! \! \! \! \overset{\tilde{\bm{r}}_{r_1} \tilde{\bm{r}}_{r_2} \tilde{\bm{r}}_{r_1} \tilde{\bm{r}}_{r_2} \tilde{\bm{r}}_{r_1} }
 {
\bcontraction{ \mathlarger{\bullet} \! \! \!- \! \! \!- \! \!}{\mathlarger{\bullet}} { \! \! \!- \! \! \!- \! \!\mathlarger{\bullet}  \! \! \!- \! \! \!- \! \!}{\mathlarger{\bullet}}
 \bcontraction[2ex]{}{ \mathlarger{\bullet} }{\! \! \!- \! \! \!- \! \!\mathlarger{\bullet}  \! \! \!- \! \! \!- \! \!}{\mathlarger{\bullet}}
 \bcontraction[2ex]{\mathlarger{\bullet} \! \! \!- \! \! \!- \! \!\mathlarger{\bullet} \! \! \!- \! \! \!- \! \!
 }
 { \mathlarger{\bullet} }{\! \! \!- \! \! \!- \! \!\mathlarger{\bullet}  \! \! \!- \! \! \!- \! \!}{\mathlarger{\bullet}}
\mathlarger{\bullet} \! \! \!- \! \! \!- \! \!\mathlarger{\bullet}  \! \! \!- \! \! \!- \! \!\mathlarger{\bullet} \! \! \!- \! \! \!- \! \!\mathlarger{\bullet}  \! \! \!- \! \! \!- \! \!\mathlarger{\bullet}
 }\! \! \! \!- _{\sigma}
 + \cdots \Biggr\rangle_{a a_0} \\
 && + 
 %
%
\sum_{\substack{r_1 \ne r_2 \ne r_3 \\ r_1 \ne r_3}}^n \Biggl\langle
 \ _{\sigma'} \! - \! \! \! \! \overset{\tilde{\bm{r}}_{r_1} \tilde{\bm{r}}_{r_3} \tilde{\bm{r}}_{r_2}}
{\mathlarger{\bullet} \! \! \!- \! \! \!- \! \!\mathlarger{\bullet}  \! \! \!- \! \! \!- \! \!\mathlarger{\bullet}}  \! \! \! \!- _{\sigma}
+   
\ _{\sigma'} \! - \! \! \! \! \overset{\tilde{\bm{r}}_{r_1} \tilde{\bm{r}}_{r_2} \tilde{\bm{r}}_{r_3} \tilde{\bm{r}}_{r_1}}
 {
 \bcontraction{}{ \mathlarger{\bullet} }{\! \! \!- \! \! \!- \! \!\mathlarger{\bullet}  \! \! \!- \! \! \!- \! \!\mathlarger{\bullet} \! \! \!- \! \! \!- \! \!}{\mathlarger{\bullet}}
\mathlarger{\bullet} \! \! \!- \! \! \!- \! \!\mathlarger{\bullet}  \! \! \!- \! \! \!- \! \!\mathlarger{\bullet} \! \! \!- \! \! \!- \! \!\mathlarger{\bullet}
 }\! \! \! \!- _{\sigma}
  +
\ _{\sigma'} \! - \! \! \! \! \overset{\tilde{\bm{r}}_{r_1} \tilde{\bm{r}}_{r_3} \tilde{\bm{r}}_{r_1} \tilde{\bm{r}}_{r_2}}
 {
 \bcontraction{}{ \mathlarger{\bullet} }{\! \! \!- \! \! \!- \! \!\mathlarger{\bullet}  \! \! \!- \! \! \!- \! \!}{\mathlarger{\bullet}}
 %
\mathlarger{\bullet} \! \! \!- \! \! \!- \! \!\mathlarger{\bullet}  \! \! \!- \! \! \!- \! \!\mathlarger{\bullet} \! \! \!- \! \! \!- \! \!\mathlarger{\bullet}
 }\! \! \! \!- _{\sigma}
 +
\ _{\sigma'} \! - \! \! \! \! \overset{\tilde{\bm{r}}_{r_1} \tilde{\bm{r}}_{r_2} \tilde{\bm{r}}_{r_3} \tilde{\bm{r}}_{r_2}}
 {
 \bcontraction{ \mathlarger{\bullet} \! \! \!- \! \! \!- \! \!}{\mathlarger{\bullet}} { \! \! \!- \! \! \!- \! \!\mathlarger{\bullet}  \! \! \!- \! \! \!- \! \!}{\mathlarger{\bullet}}
\mathlarger{\bullet} \! \! \!- \! \! \!- \! \!\mathlarger{\bullet}  \! \! \!- \! \! \!- \! \!\mathlarger{\bullet} \! \! \!- \! \! \!- \! \!\mathlarger{\bullet}
 } \! \! \! \!- _{\sigma}
 + \cdots
 \Biggr\rangle_{a a_0} \\
 && 
 + 
 \sum_{\substack{r_1 \ne r_2  \ne r_3 \ne r_4 \\ r_3 \ne r_1 \ne r_4 \ne r_2}}^n
 \Biggl\langle  \ _{\sigma'} \! - \! \! \! \! \overset{\tilde{\bm{r}}_{r_1} \tilde{\bm{r}}_{r_2} \tilde{\bm{r}}_{r_3} \tilde{\bm{r}}_{r_4} }
{\mathlarger{\bullet} \! \! \!- \! \! \!- \! \!\mathlarger{\bullet}  \! \! \!- \! \! \!- \! \!\mathlarger{\bullet} \! \! \!- \! \! \!- \! \!\mathlarger{\bullet}}  \! \! \! \!- _{\sigma}
+ \cdots \Biggr\rangle_{a a_0}  + \cdots ,
\ee
\label{RelevantDiagrams}
\end{subequations}

\noindent where each circle denotes a scattering event given by the $\tilde{\bm{r}}_r$-matrix of a slice and the continuous line between two circles, corresponding to two different slices $r_i \ne r_j$, represents the diagonal matrix  whose elements are the phase factors $e^{ik_{b}\vert x_{r_{i}} - x_{r_{j}} \vert}$. Each restricted sum symbol contains the ensemble averages of different multiple scattering process between a fixed number of scatterers; for instance, the second sum symbol with $r_{1}\neq r_{2}$ contains all the multiple scattering processes between couple of slices, where any circle denotes a scattering event given by the $\tilde{\bm{r}}$-matrix of a thin slice. The number of circles within each diagram of Eq. \eqref{RelevantDiagrams}, specifies the order in multiple scattering for the channel-channel transition $a_{0}\rightarrow a$. The open  line $-_{\sigma} =e^{ i\sigma k_{a_0}x}$ at the right hand side of each diagram represents the incoming wave in the open channel $a_{0}$, while $_{\sigma'}- =e^{ -i\sigma' k_{a}x}$ corresponds to the outgoing plane wave in the open channel $a$; the first three diagrams of Eq. \eqref{RelevantDiagrams} are the statistical average of the three terms explicitly shown in Eq. \eqref{TMwithR}.

Since the scattering units are statistically independent and identically distributed, the analysis of the series expansion of Eq. \eqref{RelevantDiagrams}, can be performed systematically.  If we denote the different moments of the generalized reflection matrices $\tilde{\bm{r}}$ [see  Eqs. \eqref{GenRefMat} and \eqref{rmoments}], as
\begin{align}
\mu^{
 (q',q)}_{b_1\cdots b_q'}
 = \left\langle ({\bf{r}}_r)_{b_1b_1^{\prime}}^* \cdots ({\bf{r}}_r)_{b_{q'}b_{q'}^{\prime}}^*  ({\bf{r}}_r)_{b_{q'+1}b_{q'+1}^{\prime}} \cdots  ({\bf{r}}_r)_{b_{q}b_{q}^{\prime}}\right\rangle,
\label{rmomentsNoAp}
\end{align}
we can then rewrite Eq. \eqref{RelevantDiagrams} in the following way
\begin{subequations}
\be
&&-\frac{i}{2}\left\langle  \mathtt{T} _{aa_{0}} ^ {\sigma^{\prime} \sigma} \right\rangle_L  =  
\mu^{
 (0,1)}_{a a_0}
\left[\sum_{r_1}^n \  _{\sigma'} \! \underset{a}{-} \! \! \overset{r_1}{*} \! \! \underset{a_0}{-} \! \! \ _{\sigma} \right]_1 
\\
&& +  \biggl\lbrace \sum_{b_1=1}^\infty  \mu^{
 (0,1)}_{a b_1}
\mu^{
 (0,1)}_{b_1 a_0}
\left[\sum_{r_1 \ne r_2}^n \! \!  _{\sigma'} \! \underset{a}{-} \! \! \overset{r_1}{*}  \! \! \underset{b_1}{-} \! \! \overset{r_2}{*} \! \! \underset{a_0}{-} \! \! \ _{\sigma} \right]_2
+ \sum_{b_1,b_2}^\infty  \! \mu^{
 (0,2)}_{a b_1 b_2 a_0}
\mu^{
 (0,1)}_{b_1 b_2}
\left[\sum_{r_1 \ne r_2}^n
\! \!  _{\sigma'} \! \underset{a}{-} \! \! \overset{r_1}{*}  \! \! \underset{b_1}{-} \! \! \overset{r_2}{*} 
\! \! \underset{b_2}{-} \! \! \overset{r_1}{*} \! \! \underset{a_0}{-} \! \! \ _{\sigma} \right]_2
 \\
 && + \sum_{b_1,b_2,b_3}^\infty
 \mu^{(0,2)}_{a b_1 b_2 b_3}  \mu^{ (0,2)}_{b_1 b_2 b_3 a_0} \!
 \left[\sum_{\substack{r_1 \ne r_2 }}^n 
\! \!  _{\sigma'} \! \underset{a}{-} \! \! \overset{r_1}{*}  \! \! \underset{b_1}{-} \! \! \overset{r_2}{*} 
\! \! \underset{b_2}{-} \! \! \overset{r_1}{*}\! \! \underset{b_3}{-} \! \! \overset{r_2}{*} \! \! \underset{a_0}{-} \! \! \ _{\sigma} \right]_2
+
\sum_{b_1,b_2,b_3,b_4}^\infty
 \mu^{(0,3)}_{a b_1 b_2 b_3 b_4 a_0}  \mu^{ (0,2)}_{b_1 b_2 b_3 b_4} \!
 \left[\sum_{\substack{r_1 \ne r_2 }}^n 
\! \!  _{\sigma'} \! \underset{a}{-} \! \! \overset{r_1}{*}  \! \! \underset{b_1}{-} 
\! \! \overset{r_2}{*} \! \! \underset{b_2}{-}
\! \! \overset{r_1}{*} \! \! \underset{b_3}{-}
\! \! \overset{r_2}{*} \! \! \underset{b_4}{-}
\! \! \overset{r_1}{*} \! \! \underset{a_0}{-} 
\! \! \ _{\sigma} \right]_2 +\cdots \biggr \rbrace 
\nonumber
 \\
 &&  +
 \biggl \lbrace 
 \sum_{b_1,b_2}^\infty \! \mu^{(0,1)}_{a b_1} 
 \mu^{(0,1)}_{b_1 b_2}
 \mu^{ (0,1)}_{b_2 a_0} \! 
\left[\sum_{\substack{r_1 \ne r_2 \ne r_3 \\ r_1 \ne r_3}}^n
\! \!   _{\sigma'} \! \underset{a}{-} \! \! \overset{r_1}{*}  \! \! \underset{b_1}{-} \! \! \overset{r_3}{*} 
\! \! \underset{b_2}{-} \! \! \overset{r_2}{*} \! \! \underset{a_0}{-} \! \! \ _{\sigma} \right]_3
+
\sum_{b_1,b_2,b_3}^\infty \! \mu^{(0,2)}_{a b_1b_3 a_0}
\mu^{ (0,1)}_{b_1 b_2} \mu^{ (0,1)}_{b_2 b_3} \!
\left[\sum_{\substack{r_1 \ne r_2 \ne r_3 \\ r_1 \ne r_3}}^n 
\! \!  _{\sigma'} \! \underset{a}{-} \! \! \overset{r_1}{*}  \! \! \underset{b_1}{-} \! \! \overset{r_2}{*} 
\! \! \underset{b_2}{-} \! \! \overset{r_3}{*}\! \! \underset{b_3}{-} \! \! \overset{r_1}{*} \! \! \underset{a_0}{-} \! \! \ _{\sigma} \right]_3
 + \cdots \biggr \rbrace 
  \\
 && +
 %
 \biggl \lbrace \sum_{b_1,b_2,b_3}^\infty \! \mu^{
 (0,1)}_{a b_1}
\mu^{
 (0,1)}_{b_1 b_2}\mu^{
 (0,1)}_{b_2 b_3}\mu^{
 (0,1)}_{b_3 a_0} \!
\left[\sum_{\substack{r_1 \ne r_2 \ne r_3 \ne r_4 \\ r_2 \ne r_4 \ne r_1 \ne r_3}}^n 
\! \!  _{\sigma'} \! \underset{a}{-} \! \! \overset{r_1}{*}  \! \! \underset{b_1}{-} \! \! \overset{r_2}{*} 
\! \! \underset{b_2}{-} \! \! \overset{r_3}{*}\! \! \underset{b_3}{-} \! \! \overset{r_4}{*} \! \! \underset{a_0}{-} \! \! \ _{\sigma} \right]_4
 + \cdots \biggr \rbrace .
\ee
\label{RelevantDiagrams2}
\end{subequations}

\noindent In Eq. \eqref{RelevantDiagrams2}, the quantities between square brackets, $[\cdots]_s$, contain the phases involved in a  multiple scattering process between $s$ slices; for instance,
\be
\left[\sum_{\substack{r_1 \ne r_2 \ne r_3 \\ r_1 \ne r_3}}^n \! \! _{\sigma'} \! \underset{a}{-} \! \! \overset{r_1}{*}  \! \! \underset{b_1}{-} \! \! \overset{r_2}{*} 
\! \! \underset{b_2}{-} \! \! \overset{r_3}{*}\! \! \underset{b_3}{-} \! \! \overset{r_1}{*} \! \! \underset{a_0}{-} \! \! \ _{\sigma}\right]_3
= \sum_{\substack{r_1 \ne r_2 \ne r_3 \\ r_1 \ne r_3}}^n e^{-i\sigma' k_a x_{r_1}} \ e^{i k_{b_1} |x_{r_1}-x_{r_2}|+i k_{b_2} |x_{r_2}-x_{r_3}|+i k_{b_3} |x_{r_3}-x_{r_1}| } \  e^{i\sigma k_{a_0} x_{r_1}},
\ee
\end{widetext}
involves 4 scattering processes between $s=3$ scattering units. The resulting expressions given in Eqs. \eqref{RelevantDiagrams2} shows  that the series expansion of the first moment $\left\langle \mathtt{T}^{\sigma' \sigma}_{aa_{0}} \right\rangle_{L}$, depends on the microscopic statistics through the statistical moments of the generalized reflection matrix of the individual slices, $\mu^{(0,q)}_{b_1\cdots b_q'}$, where the channel indexes could be any possible combination of open and closed channels, so any possible channel-channel transition is captured in this perturvative approach.

\subsection{\label{TheNewDWSL}The limiting macroscopic statistics of the Building Block}

In order to obtain the limiting macroscopic statistic of the BB, it is convenient to use a simplified mathematical limit, where the BB can be considered as a continuous system. For that purpose, we use a generalization of the so called {\em dense weak scattering limit} (DWSL), which has been used to obtain limiting macroscopic statistics in previous studies  \cite{AnnaPhys.181.290,PhysRevB.37.5860,PhysRevB.46.15963,PhysRevE.75.031113,EPL.108.17006,*YepezSaenz:arXiv:1407.5617}. In the DWSL, the individual scattering units are considered extremely weak, the number of scatterers $n$ goes to infinity, the distance between consecutive slices $d=f^{-1} \delta$ tends to zero (dense or continuous limit), but the length $L$ of the BB remains fixed, i.e.,
\begin{subequations}
\begin{eqnarray}
d= \frac{\delta}{f}  \rightarrow 0,
\quad 
n \rightarrow \infty ,
\;\;\;
L=n \frac{\delta}{f}  \;\;\; \mathrm{fixed}.
\label{dense_limite}
\end{eqnarray}
In addition, the different moments of the generalized reflection matrices $\tilde{\bm{r}}$ [see  Eqs. \eqref{GenRefMat} and \eqref{rmoments}], $\mu^{(q',q)}_{b_1\cdots b_q'}$, obey the  scaling law
\begin{eqnarray}
\lim_{ \delta \rightarrow 0}  \ \ \ && \frac{1}{\delta^p}\mu^{
 (0,2p)}_{b_1\cdots b_{2p}'} = \alpha_{b_1b_1' \cdots b_{2p}b_{2p}^{\prime}}^{(2p)}  \;\; \mathrm{finite} \\
\lim_{ \delta \rightarrow 0}  \ \ \ && \frac{1}{\delta^p}\mu^{ (0,2p-1)}_{b_1\cdots b_{2p-1}'}  = \alpha_{b_1b_1' \cdots b_{2p-1}b_{2p-1}^{\prime}}^{(2p-1)} 
\;\; \mathrm{finite}
\label{Weaklimit}
\end{eqnarray}
\label{Weak_dense_limits}
\end{subequations}

\noindent as the slice thickness goes to zero [see Eq. \eqref{scalingmu}].

We can now proceed to analyse the series expansion of Eq. \eqref{RelevantDiagrams2} in the DWSL defined by Eqs. \eqref{Weak_dense_limits}. We first notice that in the dense limit, Eq. \eqref{dense_limite}, the restricted sums over scattering units in Eq.  \eqref{RelevantDiagrams2} transform into integrals
\be
\sum_{r_1=1}^n &\rightarrow & \frac{f}{\delta} \int_0^L d x_{1},
\nonumber
\\
\sum_{r_1 \ne r_2}^n &\rightarrow &\frac{f^2}{\delta^{2}} \iint_0^L d x_{1} d x_{2}, \nonumber
\\&\vdots  & ;
\ee
therefore, each square bracket quantity, $[\cdots]_s$ in Eq. \eqref{RelevantDiagrams2} (involving $s$ different slices) scales as 
\be
[\cdots]_s \overset{\mathrm{DWSL}}{\sim} \frac{1}{\delta^s} \label{squarescale}
\ee
Second, we also notice that each term involving $s$ scattering units contains the product of $s$ different moments
\be
\left\{\mu^{
 (0,q_1)}_{\cdots}\mu^{
 (0,q_2)}_{\cdots} \cdots \mu^{
 (0,q_s)}_{\cdots}\right\} \big[\cdots\big]_s.
\ee
From Eqs. \eqref{Weak_dense_limits} and \eqref{squarescale}, it is easy to conclude that, in the DWSL, all those terms containing $q_i$-moments higher than 2 do not contribute to the expansion of $\left\langle \mathtt{T}^{\sigma' \sigma}_{aa_{0}} \right\rangle_{L}$, i.e.,
\be
\lim_{ \delta \rightarrow 0}  \ \ \ \left\{\mu^{
 (0,q_1)}_{\cdots}\mu^{
 (0,q_2)}_{\cdots} \cdots \mu^{
 (0,q_s)}_{\cdots}\right\} \big[\cdots\big]_s = 0 \nonumber \\  \mathrm{if} \ q_1 >2, \; \mathrm{or} \; q_2 >2,  \; \cdots, \; \mathrm{or} \; q_s >2 \nonumber 
\ee
The above analysis demonstrates that, in the DWSL, the expectation value $\left\langle  \mathtt{T}^{\sigma' \sigma}_{aa_{0}} \right \rangle_{L}$ of the BB, depend only on the microscopic details through the first and second moments of the slices
\begin{subequations}
\begin{eqnarray}
f \alpha_{bb^{\prime}}^{\left(1\right)}&=& \lim_{DWS} \frac{\mu_{bb^{\prime}}^{(0,1)}}{d} =
\lim_{DWS}\frac{\left\langle \left(\widetilde{\bm{r}}_{r}\right)_{bb^{\prime}} \right\rangle}{d},
\label{Relevant_1st_Moment}
\\
f \alpha _{bb^{\prime} cc^{\prime}} ^{\left(2\right)}&=&
\lim_{DWS}\frac{\mu_{bb^{\prime} cc^{\prime}}^{(0,2)}}{d} =
\lim_{DWS}\frac{
\left\langle
\left(\widetilde{\bm{r}}_{r}\right)_{bb^{\prime}}
\left(\widetilde{\bm{r}}_{r}\right)_{cc^{\prime}}
\right\rangle}{d},
\label{Relevant_2nd_Momentbis}
\ee
\label{Relevan_Statistical_moments}
\end{subequations}

\noindent where $f \alpha_{bb^{\prime}}^{\left(1\right)}$ and  $f \alpha _{bb^{\prime} cc^{\prime}} ^{\left(2\right)}$ have the units of inverse lengths. 

Second moments of the transition matrix elements, $\left\langle \vert \mathtt{T}^{\sigma' \sigma}_{aa_{0}}  \vert^{2} \right \rangle_{L}$, as well as higher moments, depend on the microscopic statistics through $\mu^{(q',q)}_{b_1\cdots b_q'}$, which also involves complex conjugate matrix elements. A completely similar procedure allows us to find the additional set of  second order moments
\be
f \eta _{bb^{\prime} cc^{\prime}} ^{\left(1,2\right)}&=&
\lim_{DWS}\frac{\mu_{bb^{\prime} cc^{\prime}}^{(1,2)}}{d} =
\lim_{DWS}\frac{
\left\langle
\left(\widetilde{\bm{r}}_{1}\right)_{bb^{\prime}}^{\ast}
\left(\widetilde{\bm{r}}_{1}\right)_{cc^{\prime}}
\right\rangle}{d},
\label{Relevant_2nd_Moment}
\end{eqnarray}
\noindent while higher moments do not play any role in the limiting macroscopic statistics. However, as it is shown in Appendix \ref{StatMomrAp} [see Eqs. \eqref{rrsinconj} and \eqref{rrconj}], in the limit of very thin slices, $\eta _{bb^{\prime} cc^{\prime}} ^{\left(1,2\right)} = -\alpha _{bb^{\prime} cc^{\prime}} ^{\left(2\right)}$, which show that the limiting macroscopic statistics depends only on microscopic details through the first and second statistical moments $\left( \widetilde{\bm{r}}_{r}\right)_{bb^{\prime}}$, Eq. \eqref{Relevan_Statistical_moments}. 

The analysis presented here, signals the existence of a generalized {\em central-limit theorem} (CLT) for the statistical scattering properties of the BB: once the first and second moments of the generalized reflection matrix elements  are specified, the macroscopic statistics of the BB is universal and independent of other microscopic details. 

\section{\label{Gen_CLT} Scaling parameters in the short wavelength approximation}

Although the CLT shows that all microscopic details are contained in just the first and second moments of the generalized reflection matrix elements, the macroscopic statistics of the BB still depends formally on an infinite number of statistical parameters related to channel-channel transitions involving both open and (an infinite number of) closed channels indexes. This is in contrast with the rather small (finite) set of characteristic lengths, or {\em mean free paths} (mfp), $\ell$, $\ell_{aa_{0}}$, $\ell_{a_{0}}$ and $\ell_{a_{0}}'$ found in previous theoretical studies \cite{PhysRevLett_89_246403,PhysRevB.46.15963,PhysRevE.75.031113,EPL.108.17006,*YepezSaenz:arXiv:1407.5617}. However, these characteristic lengths can be identified as some of the relevant statistical moments introduced in  Eqs. \eqref{Relevan_Statistical_moments}-\eqref{Relevant_2nd_Moment}:
\begin{subequations}
\be 
-\frac{1}{\ell_{a_0}} + i \frac{1}{\ell_{a_0}'}
\equiv 
f \alpha_{a_0 a_0}^{\left(1\right)}
&=&
\lim_{DWS}\frac{\left\langle \left(\widetilde{\bm{r}}_{r}\right)_{a_0a_0} \right\rangle}{d},
\label{twolengths}
\\
\frac{1}{\ell_{a a_0}}
\equiv 
f \eta _{a a_0 a a_0} ^{\left(1,2\right)} 
&=&
\lim_{DWS}\frac{
\left\langle
\left(\widetilde{\bm{r}}_{1}\right)_{aa_0}^{\ast}
\left(\widetilde{\bm{r}}_{1}\right)_{aa_0}
\right\rangle}{d},
\\
\equiv
- f \alpha _{a a_0 a a_0} ^{\left(2\right)}
&=&
-\lim_{DWS}\frac{
\left\langle
\left(\widetilde{\bm{r}}_{1}\right)_{aa_0}
\left(\widetilde{\bm{r}}_{1}\right)_{aa_0}
\right\rangle}{d}.
\nonumber
\ee
\end{subequations}

In order to understand this difference, we notice that the derivations of previous theoretical works, consider physical situations in which the incident wavelength $\lambda$ is much smaller than both, the elastic mean free path
\be
\frac{1}{\ell} &=&\frac{1}{N} \sum_{a,a_{0}=1}^N \frac{1}{\ell_{aa_0}},
\ee
and the length of the system, i.e., $\lambda \ll  \ell,L$. If we denote symbolically any characteristic length $\ell_{aa_{0}}$, $\ell_{a_{0}}$, $\ell_{a_{0}}^{\prime}$ as $\ell$ and any real or imaginary longitudinal wave number $k_{b}$ simply by  $k$, the relation
\begin{equation}  
k\ell\gg 1, \quad kL \gg 1,
\label{MyPHDTEq4.43} 
\end{equation}
defines the so called {\em short-wavelength approximation} (SWLA), because its similarity to the geometrical optics limit \cite{PhysRevE.75.031113,BornandWolf:1999}. The approximation $k\ell\gg 1$, which is best known in the literature as the weak disorder condition \cite{RevModPhys.69.731}, has been used widely in the derivation of previous theoretical approaches for the wave transport in disordered systems.

Let us consider again the square brackets, $[\cdots]_s$ in Eq. \eqref{RelevantDiagrams2}. It is easy to see that any multiple scattering contribution to the  average of the channel-channel $a_0 \rightarrow a$ transition matrix elements that take place through any intermediate evanescent channel $b_1>N$ (i.e., $k_{b_1} = i\kappa_{b_1}$), contains an exponential factor $e^{-\kappa_{b_1}|x_{r_i}-x_{r_j}|}$ ($x_{r_i} \ne x_{r_j}$), which, in the DWSL, give rise to contributions of the order or smaller than $e^{-\kappa_{b_1}L}$; therefore, these terms can be neglected in the SWLA, Eq. \eqref{MyPHDTEq4.43}. The same applies for higher moments of the transition matrix elements. This is valid as long as the wave number $k$ is far from a mode threshold, where $\kappa_{b_1}=k\sqrt{(\pi b_{1}/kW)^{2}-1)}$ can be $\sim 0$ when $b_1 = N+1$ is the first evanescent mode; therefore, transport through evanescent modes can be neglected in the SWLA provided $N \lnsim kW/\pi \lnsim N+1$  (see App. \ref{StatMomrAp}).

As an example of an application of the SWLA, we explicitly compute the {\em the dominant contributions} [of the order $\sim 1/(k\ell)^{0}$] to the averaged transmission amplitude [see Appendix \ref{Avtra} for details]
\be
\left\langle t_{a a_{0}}\right\rangle _{L}  &=& \delta_{aa_0} \left\{1+f\alpha_{a_0a_{0}}^{\left(1\right)}  L + \frac{1}{2!}(f \alpha_{a_{0}a_0} ^{\left(1\right)} L )^2 + \cdots \right\} \nonumber \\ &=& \delta_{aa_0} e^{f\alpha_{a_0a_{0}}^{\left(1\right)}  L}= \delta_{aa_0}e^{iL/\ell_{a_0}'}e^{-L/\ell_{a_0}}.
\label{AverTransAmplitude}
\ee
In this simple case, the only characteristic lengths that appears are $\ell_{a_0}$ and $\ell_{a_0}'$. This result is consistent with a previous work based on the Born series expansion discussed in Ref. \cite{EPL.108.17006,*YepezSaenz:arXiv:1407.5617}, where $\ell_{a_0}'$ and $\ell_{a_0}$ are connected with the effective wave number in disordered waveguides. In App. \ref{OTSS}, it is demonstrated that the scattering MFP $\ell_{a_0}$ and the channel-channel MFP's $\ell_{a a_0}$ are related by
\be
\frac{1}{\ell_{a_0}} &=& 
- \lim_{DWS}\frac{\left\langle \text{Re}\left\{ \left(\widetilde{\bm{r}}_{r}\right)_{a_0a_0} \right\}\right\rangle}{d}
\nonumber
\\
&=& \lim_{DWS}
\sum_{a=1}^{N}\frac{
\left\langle
\vert
\left(\widetilde{\bm{r}}_{r}\right)_{aa_0}
\vert^{2}
\right\rangle}{d}
=\sum_{a=1}^{N} \frac{1}{\ell_{aa_0}},
\label{TwoDefs_ScatMFP}
\ee
\noindent which is consequence of the Optical Theorem for a single slice, Eq. \eqref{OptTheoSlice}. Equation \eqref{TwoDefs_ScatMFP} shows that it is not possible to impose arbitrary assumptions on the statistical scattering properties of the individual scatters. For instance, in the well known {\em isotropic distribution model of phases} (IM) \cite{JMP.27.2876,PhysRevB.37.5860,PhysRevB.44.3559}, the complex phases of the polar representation of the scattering matrix \cite{Mello:2010}, are assumed statistically independent and uniformly distributed, which leads to the following expressions

\begin{equation}
\left. \langle (\widetilde{\bm{r}}_{r} )_{a_{0}a_{0}} \right. \rangle ^{ \mathrm{( IM)} } =0, \;\;\; \left. \langle \vert ( \widetilde{\bm{r}}_{r} )_{aa_{0}} \vert^{2} \right. \rangle ^{\mathrm{( IM)} } \neq 0;
\label{IM_relations}
\end{equation}

\noindent here \begin{small}$\langle\vert(\widetilde{\bm{r}}_{r})_{aa_{0}}\vert^{2}\rangle^{\mathrm{( IM)} }$\end{small} and \begin{small}$\langle(\widetilde{\bm{r}}_{r})_{a_{0}a_{0}}\rangle ^{\mathrm{( IM)} }$\end{small} denote, respectively, the statistical averages in the IM model of the reflection coefficient and reflection amplitude of a single slice. However, the IM model expressions of Eq. \eqref{IM_relations}, are not consistent with the OT constriction of Eq. \eqref{TwoDefs_ScatMFP}.

\noindent Even though the optical theorem does not give any information about \begin{small}$ \mathrm{Im}\left. \langle (\tilde{ \bm{r}}_{r} )_{a_{0}a_{0}} \right. \rangle $\end{small}, the length scale
\begin{equation}
\frac{1}{\ell_{a_{0}}^{\prime}}
\equiv
\lim_{DWS}\frac{\left\langle \text{Im}\left\{ \left(\widetilde{\bm{r}}_{r}\right)_{a_0a_0} \right\}\right\rangle}{d},
\label{NonOpTheAnalgous}
\end{equation}
was introduced in Ref. \cite{EPL.108.17006,*YepezSaenz:arXiv:1407.5617}, which captures the closed channel influence on the statistics of the scattering amplitudes of a disordered waveguide.

\section{\label{Conclusions}Conclusions}

In the present work we demonstrate that the limiting macroscopic statistics of disordered waveguides emerges as consequence of a generalized Central Limit Theorem, in the sense that the expectation values of macroscopic observables depend only on the microscopic details of the disorder through the first and second moments of the generalized reflection matrix elements of the individual scatterers. In contrast with previous approaches, the CLT found in the present work is consistent with the optical theorem. This is an important fundamental issue since the well known hypothesis of isotropic distribution of phases, used as the starting point of earlier works, does not satisfy the optical theorem relation. This assumption imposes a null value for the scattering amplitude in the forward direction, while the average of the total scattered flux is not zero. In addition, the optical theorem also showed the way forward to introduce the new characteristic length $\ell_{a_{0}}^{\prime}$, Eq. \eqref{NonOpTheAnalgous}, that captures the closed channel contributions in the macroscopic statistic of disordered waveguides.
 
The generalized CLT represents some other important advantages compared with previous ones: i) The energy of the incident particle is fully taken into account. ii) This CLT allows us to identify the order in multiple scattering of the wave transport, where open and closed channels transitions are included through the statistical moments of Eqs. \eqref{Relevan_Statistical_moments}-\eqref{Relevant_2nd_Moment}. iii) The derivation of the present CLT, does not contain any assumption on the statistical scattering properties of the BB, or on the statistic of the individual scattering units.

\section*{Acknowledgments}

This work was supported by the  Spanish Ministry of Economy and Competitiveness  (grant number FIS2012-36113-C03 ). MY thanks the Mexican CONACyT for the postdoctoral grants No. 162768, 187138.

\appendix

\section{\label{Tmss}Scattering and statistical properties of a thin slice}

\subsection{\label{OTSS} Transition matrix and the Optical Theorem for a thin slice}

Let us consider the simple case of a single slice.
Since $\lambda \gg \delta$, the wave function can be assumed to be constant {\em inside} the slice along the wave propagation direction $x$, i.e.
\begin{equation}
\Psi_{\sigma a_{0}}\left(x, y\right) = \sum_{b=1}^{\infty} \left[ \psi_{\sigma a_{0}}\left(x_r\right) \right]_{b} \chi_{b}\left( y \right), \quad |x-x_r| < \frac{\delta}{2}.
\end{equation}
Evaluating Eq. \eqref{MY_PHDT_2_28ab} {\em inside} the slice, we obtain
\begin{eqnarray}
\left[ \psi _{\sigma a_{0}}\left(x_r\right) \right]_{b}&=&\varphi _{\sigma}\left( k_{b};x_r\right) \delta_{ba_{0}}
\label{MY_PHDT_2_28absingle}
\\
&-&\sum_{b_{1}=1}^{\infty} [\widehat{\bm{G}}_0]_{bb}
\left[({\bf{u}}_r)_{bb_{1}} \delta\right] \left[ \psi _{\sigma a_{0}}\left(x_r\right) \right]_{b_{1}},
\nonumber
\end{eqnarray}
where we have defined
\begin{equation}
[\widehat{\bm{G}}_0]_{bb}\equiv \frac{1}{\delta}  \int_{x_{r}-\delta/2}^{x_{r}+\delta/2} [G_0(x_r,x)]_{bb} dx
= \frac{i }{2 k_{b}}
\left( 
\frac{e^{i\frac{ k_{b} \delta}{2}}-1}{\frac{i k_{b}\delta}{2}}
\right),
\label{SpaceAverageGreenFunction}
\end{equation}
as the average (over the slice) of the Green function of channel $b$ [see Eq. \eqref{GFb}]. The self-consistent set of equations \eqref{MY_PHDT_2_28absingle} can be written in compact form 
\be
\left[ \bm{\psi} _{\sigma}\left(x_r\right) \right] &=&\bm{\varphi} _{\sigma}\left(x_r\right) 
- \widehat{\bm{G}}_0
{\bf{u}}_r\delta  \left[ \bm{\psi} _{\sigma} \left(x_r\right)\right] ,
\ee
whose solution is given by 
\be
\left[ \bm{\psi} _{\sigma}\left(x_{r}\right)\right] = \left[\frac{\bf{I}}{\bf{I}+\widehat{\bf{G}}_0 \bf{u}_{r} \delta} \right] \bm{\varphi}_{\sigma}\left(x_{r}\right),
\label{SolutionSingleSlice}
\ee
$\bf{I}$ being the identity matrix.

By substituting Eq. \eqref{SolutionSingleSlice} into the Eq. \eqref{MY_PHDT_2_28ab} for a single slice, it is easy to find the wave function outside the scatter
\begin{eqnarray}
\left[ \psi _{\sigma a_{0}}\left(x\right) \right]_{b}&=&\varphi _{\sigma}\left( k_{b};x\right) \delta_{ba_{0}}
\label{WaveFunctioOutScatter}
\\
&+& g_{0}\left(k_{b};x,x_r \right) \left[\bm{\mathcal{T}}_r \right]_{b a_0} \varphi_{\sigma}\left(k_{a_{0}};x_{r}\right),
\nonumber
\end{eqnarray}
where we have defined the $\bm{\mathcal{T}}_r$-matrix as
\be
\bm{\mathcal{T}}_r &=& \left[ \bf{u}_r \delta
\frac{\bf{I}}{\bf{I}+\widehat{\bf{G}}_0 \bf{u}_{r} \delta} \right];
\label{Tmatrix1S}
\ee
notice that in Eq. \eqref{WaveFunctioOutScatter}, channel $b$ can be either an open (propagating mode) or a closed (evanescent mode) channel, as discussed in Ref. \cite{EPL.108.17006,*YepezSaenz:arXiv:1407.5617}. The $\bm{\mathcal{T}}_r$-matrix defines the (dimensionless) generalised reflection matrix $\tilde{\bm{r}}_r$ of a single slice as
\be
\tilde{\bm{r}}_r &\equiv&  - \frac{i}{2} \frac{1}{\sqrt{\bm k}} \bm{\mathcal{T}}_r  \frac{1}{\sqrt{\bm k}},
\label{grm}
\\
(\tilde{\bm{r}}_r)_{bb'} &\equiv& - \frac{i}{2} \frac{1}{\sqrt{k_b}} \left[ \bm{\mathcal{T}}_r \right]_{bb'} \frac{1}{\sqrt{k_{b^{\prime}}}}.
\label{GenRefMat}
\ee

From Eqs. \eqref{TMT} and \eqref{ReDef_Back_For_scatt_Ampli}, we obtain the transition matrix operator for a subwavelength slice,
\be
\widehat{ \mathbb{T} }_r \left( x^{\prime },x^{\prime \prime}\right) = \delta(x'-x_r) \ \bm{\mathcal{T}}_r \ \delta(x''-x_r),
\label{Toperator1S}
\ee
and the corresponding transition matrices
\begin{subequations}
\begin{eqnarray}
\mathtt{T} _{aa_{0}} ^ {\sigma^{\prime} \sigma} &=& \varphi_{\sigma^{\prime}} ^{\ast} \left(k_{a};x_r\right) [\bm{\mathcal{T}}_r ]_{aa_{0}}  \varphi_{\sigma} \left(k_{a_{0}};x_r\right)
\\
&=&2i [\tilde{\bm{r}}_r ]_{aa_{0}}  \ e^{i(\sigma k_{a_0}-\sigma' k_a) x_r}.
\end{eqnarray}
The reflection and trasmission amplitudes follow directly from Eqs. \eqref{ScattAmplandTMatrix}
\be
r_{aa_{0}}&=&(\tilde{\bm{r}}_r)_{aa_{0}} e^{i(k_{a_0}+ k_a) x_r},
\\
t_{aa_{0}}&=&\delta_{aa_{0}} + (\tilde{\bm{r}}_r)_{aa_{0}} e^{i(k_{a_0}- k_a) x_r}.
\ee
\label{ScattAmplandTMatrixSlice}
\end{subequations}

It is easy to check that the above relations fulfil the  Optical Theorem,  Eq. \eqref{GeneralOpticalTheorem}, for a single slice: 
\be
\sum_{a=1}^N  
\vert \left(\widetilde{\bm{r}}_{r}\right)_{aa_0}\vert ^{2}
=- \text{Re} \left\{ \left(\widetilde{\bm{r}}_{r}\right)_{a_0a_0} \right\},
\label{OptTheoSlice}
\ee
and impose the following statistical constrictions for the statistical averages

\begin{subequations}
\begin{eqnarray}
&&\sum_{a=1}^{N} \left. \langle \vert (\widetilde{\bm{r}}_{r} )_{aa_{0}}
\vert^{2} \right. \rangle =-\mathrm{Re} \left. \langle ( \widetilde{\bm{r}}_{r} )_{a_{0}a_{0}} \right. \rangle,
\label{OpTheConstriction}
\\
&&\sum_{a=1}^{N}\frac{\left \langle \vert \left( \bm{\mathcal{T}}_{r}\right)_{aa_{0}}\vert^{2}\right\rangle}{4k_{a}k_{a_{0}}}=-\frac{\mathrm{Im}
\left\langle \left(\bm{ \mathcal{T}}_{r}\right)_{a_{0}a_{0}} \right\rangle 
}{2k_{a_{0}}}.
\label{OT_Single_Slice}
\end{eqnarray}
\label{FCandOT_slice}
\end{subequations}

\subsection{\label{StatMomrAp} Statistical moments of the $\tilde{\bm{r}}$-matrix}

In the weak scattering limit \cite{EPL.108.17006,*YepezSaenz:arXiv:1407.5617}, $\tilde{\bm{r}}_r$ can be expanded in a Born series
\be
{\tilde{\bm{r}}}_r =  -\frac{i}{2} \delta \frac{1}{\sqrt{{\bm{ k}}}} {\bf{u}}_r   \frac{1}{\sqrt{{\bm {k}}}} + \frac{i}{2}\delta^2  \frac{1}{\sqrt{\bm k}} \bf{u}_r  \widehat{\bf{G}} \bf{u}_{r}  \frac{1}{\sqrt{\bm k}} + \cdots .
\nonumber
\\
\label{Bornr}
\ee
This expansion is valid as long as any wave number $k_b$ is far away from the threshold of the last propagating mode and the first evanescent mode. This avoids the resonant phenomena arising from the strong coupling to the first evanescent mode, that appears when $k_b \sim 0$ near to the threshold of a new propagating mode \cite{PhysRevB.41.10354,*kunze1992single,*PhysRevB.50.17415,*gomez2001resonant}.

Let us define $\mu^{
 (q',q)}_{b_1\cdots b_q'}$ as the moments of order $q$ involving the products of $(q-q')$  matrix elements and $q'$ complex conjugates, i.e.
\begin{align}
\mu^{
 (q',q)}_{b_1\cdots b_q'}
 = \left\langle ({\bf{r}}_r)_{b_1b_1^{\prime}}^* \cdots ({\bf{r}}_r)_{b_{q'}b_{q'}^{\prime}}^*  ({\bf{r}}_r)_{b_{q'+1}b_{q'+1}^{\prime}} \cdots  ({\bf{r}}_r)_{b_{q}b_{q}^{\prime}}        \right\rangle. \label{rmoments}
 \end{align}
In the limit of very thin slices ($\delta \rightarrow 0$),   the statistical moments of the (complex) $\tilde{\bm{r}}_r$ elements [that can be obtained from Eq. \eqref{Bornr} together with Eqs. \eqref{oddmomentsu} and \eqref{evenmomentsu}]
are found to obey the following scaling laws with $\delta$:
\be
\left\langle \left[\tilde{\bm{r}}_r \right]_{b_1b_1^{\prime}}\right\rangle &\overset{\delta \rightarrow 0}{\sim}&
\left\{\frac{i}{2 } \sum_{b_2=1}^\infty \frac{\mu^{(\tilde{\bf{u}}_r)}_{b_1 b_2 b_2  b_{1}'} [\widehat{\bf{G}}]_{b_2 b_2}}{\sqrt{k_{b_1}k_{b_1'}}} \right\} \delta + \mathcal{O}
\left\{\delta^3\right\}  \nonumber \\ &=&
 \alpha_{b_1b_1'}^{(1)}  \delta + \mathcal{O}
\left\{\delta^3\right\}  
\ee
\be
\left\langle \left[\tilde{\bm{r}}_r \right]_{b_1b_1^{\prime}} \left[\tilde{\bm{r}}_r \right]_{b_2b_2^{\prime}}   \right\rangle &\overset{\delta \rightarrow 0}{\sim}&
-\left\{\frac{1}{4 } \frac{\mu^{(\tilde{\bf{u}}_r)}_{b_1 b_1' b_2  b_{2}'}}{ \sqrt{ k_{b_1}k_{b_1'} k_{b_2}k_{b_2'}  }  } \right\} \delta + \mathcal{O}
\left\{\delta^3\right\}  \nonumber \\ &=&
 \alpha_{b_1 b_1' b_2  b_{2}'}^{(2)}  \delta + \mathcal{O}
\left\{\delta^3\right\}  \label{rrsinconj}
\\
\left\langle \left[\tilde{\bm{r}}_r \right]_{b_1b_1^{\prime}} \left[\tilde{\bm{r}}_r \right]_{b_2b_2^{\prime}}^*   \right\rangle &\overset{\delta \rightarrow 0}{\sim}&
+\left\{\frac{1}{4 } \frac{\mu^{(\tilde{\bf{u}}_r)}_{b_1 b_1' b_2  b_{2}'}}{ \sqrt{ k_{b_1}k_{b_1'} k_{b_2}k_{b_2'}  }  } \right\} \delta + \mathcal{O}
\left\{\delta^3\right\}  \nonumber \\ &=&
 \eta_{b_1 b_1' b_2  b_{2}'}^{(1,2)}  \delta + \mathcal{O}
\left\{\delta^3\right\}   \nonumber \\ &=&
- \alpha_{b_1 b_1' b_2  b_{2}'}^{(2)}  \delta + \mathcal{O}
\left\{\delta^3\right\}  \label{rrconj}
\\
&\cdots& \nonumber 
\ee
\be
\left.
\begin{array}{lcr}
\mu^{
 (0,2p-1)}_{b_1\cdots b_{2p-1}'}
&\overset{\delta \rightarrow 0}{\sim}& \alpha_{b_1b_1' \cdots b_{2p-1}b_{2p-1}^{\prime}}^{(2p-1)}  \delta^{p} + \mathcal{O} 
\left\{\delta^{p+2}\right\}   \\
\mu^{
 (q',2p-1)}_{b_1\cdots b_{2p-1}'}
&\overset{\delta \rightarrow 0}{\sim}& \eta_{b_1b_1' \cdots b_{2p-1}b_{2p-1}^{\prime}}^{(q,2p-1)}  \delta^{p} + \mathcal{O} 
\left\{\delta^{p+2}\right\}   \\ & & \{1 \le q \le 2p-1 \} \\
\end{array}
\right\rbrace
\text{odd} \nonumber
\\ 
\left.
\begin{array}{lcr}
\mu^{(0,2p)}_{b_1\cdots b_{2p-1}'}
&\overset{\delta \rightarrow 0}{\sim}& \alpha_{b_1b_1' \cdots b_{2p}b_{2p}^{\prime}}^{(2p)}  \delta^{p} + \mathcal{O} 
\left\{\delta^{p+2}\right\}
\\
\mu^{(q',2p)}_{b_1\cdots b_{2p}'}
&\overset{\delta \rightarrow 0}{\sim}& \eta_{b_1b_1' \cdots b_{2p}b_{2p}^{\prime}}^{(q,2p)}  \delta^{p} + \mathcal{O} 
\left\{\delta^{p+2}\right\}    \\ & &  \{1 \le q \le 2p \}
\\
\end{array}
\right\rbrace
\text{even} \nonumber \\
\label{scalingmu}
\ee
where $\alpha _{\cdots}^{\left( \cdots\right) } $ and $\eta _{\cdots}^{\left( \cdots\right) } $ are, in general, complex quantities that depend on $k$, but do not depend on $\delta$. 

In summary, $2p$-moments and $(2p-1)$-moments of the generalized reflection matrix elements  scale with the slice thickness as $\delta^p$.

\section{\label{FinalGenScattTheory}Perturbative expansion}

The set of Lippmann-Schwinger coupled equations [Eqs. \eqref{MY_PHDT_2_28ab}, \eqref{CompactLimppman} and \eqref{TMT}] for our set of $n$ scattering units (slices) can be written in compact form as
\be
\bm{\psi}_{\sigma}\left(x\right) 
&=&
\bm{\varphi}_{\sigma}\left( x\right) 
-\sum_{r_{1}=1}^n  \bm{G}_0\left(x,x_{r_{1}}\right)  \left[ { \bm{u}}_{r_{1}} \delta \
\bm{\psi}_{\sigma } \left(x_{r_{1}}\right) \right],
\label{MY_PHDT_2_28abAp}
\ee
where the wave function inside each slice has been considered constant. The Lippmann-Schwinger equations, Eq. \eqref{CompactLimppman}, allows us to evaluate the solution inside the slice located at $x=x_{r_1}$ as
\be
\bm{\psi}_{\sigma } \left(x_{r_1}\right)
&=&
\bm{\varphi}_{\sigma} \left(x_{r_{1}} \right)
-\widehat{\bf{G}}_0  \left[ { \bm{u}}_{r_1}\delta \
\bm{\psi}_{\sigma } \left(x_{r_1}\right) \right]
\nonumber
\\
&-& \sum_{\substack{r_2 \\ r_2 \ne r_1}}^n  \bm{G}_0 \left(x_{r_1},x_{r_2}\right) \left[ { \bm{u}}_{r_2} \delta \
\bm{\psi}_{\sigma } \left(x_{r_2}\right) \right];
\ee
notice that we have separated the {\em self interaction} contributions of the scatterer centered at $x_{r_{1}}$ (second term on the right hand side), from those interactions coming from rest of the scatterers $x_{r_{2}} \neq x_{r_{1}}$ (third term on the right hand side). The above equation gives rise to the following relation
\be
\left[ { \bm{u}}_{r_1} \delta \
\bm{\psi}_{\sigma } \left(x_{r_1}\right) \right]
&=& \bm{\mathcal{T}}_{r_1} \bm{\varphi}_{\sigma } \left(x_{r_1}\right)
\label{Recursion}
\\
-&& \bm{\mathcal{T}}_{r_1} \sum_{\substack{r_2 = 1\\ r_2 \ne r_1}}^n  \bm{G}_0 \left(x_{r_1},x_{r_2}\right) \left[ { \bm{u}}_{r_2} \delta \
\bm{\psi}_{\sigma } \left(x_{r_2}\right) \right] \nonumber
\ee
where the $\bm{\mathcal{T}}$-matrix of the slice $r_1$ has been used: see Eq. \eqref{Tmatrix1S}. In analogy with the standard perturbation theory of Ref. \cite{AnndAstPhy.30.565}, the iteration of the recursion relation of Eq. \eqref{Recursion}, allows us to write the transition matrix operator $\widehat{ \mathbb{T} } \left( x,x'\right)$, Eq. \eqref{TMT}, into a series expansion in powers of the individual $\widehat{ \mathbb{T} }_r$ of the single slices, Eq. \eqref{Toperator1S}, i.e., 
\be
&& \widehat{ \mathbb{T} } \left( x,x'\right)
=  \sum_{r_1=1}^n \widehat{ \mathbb{T} }_{r_1}\left( x,x'\right)
\label{1rstIteration}
\\
-&&\sum_{\substack{r_1,r_2 \\ r_2 \ne r_1}}^n  \iint_0^L \widehat{ \mathbb{T} }_{r_1}\left( x,x''\right)   \bm{G}_0 \left(x'',x'''\right) \widehat{ \mathbb{T} }_{r_2}\left( x''',x'\right) dx'' dx'''
\nonumber
\\
+&& \cdots, \nonumber
\ee
which can be formally written as a diagrammatic expansion:
\begin{widetext}
\be
\MT &=&  \sum_{r_1}^n \overset{\MT_{r_1} }
{\mathlarger{\bullet} }
+ \sum_{r_1 \ne r_2}^n \ \ \overset{\MT_{r_1} \MT_{r_2} }
{\mathlarger{\bullet} \! \! \!- \! \! \!- \! \!\mathlarger{\bullet} } 
+ 
\Biggr\{ 
\sum_{r_1 \ne r_2}^n 
 \overset{\MT_{r_1} \MT_{r_2} \MT_{r_1}}
 {
 \bcontraction{}{ \mathlarger{\bullet} }{ \! \! \!- \! \! \!- \! \!\mathlarger{\bullet}  \! \! \!- \! \! \!- \! \!}{\mathlarger{\bullet}}
 \mathlarger{\bullet} \! \! \!- \! \! \!- \! \!\mathlarger{\bullet}  \! \! \!- \! \! \!- \! \!\mathlarger{\bullet}
 }
%
+
\sum_{\substack{r_1 \ne r_2 \ne r_3\\ r_1 \ne r_3}}^n
 \overset{\MT_{r_1} \MT_{r_2} \MT_{r_3}}
{\mathlarger{\bullet} \! \! \!- \! \! \!- \! \!\mathlarger{\bullet}  \! \! \!- \! \! \!- \! \!\mathlarger{\bullet}}  
  \Biggr\} \nonumber \\ 
+  &&   \! \! \! \! \! \! \Biggr\{ 
\sum_{r_1 \ne r_2 }^n
\overset{\MT_{r_1} \MT_{r_2} \MT_{r_1} \MT_{r_2}}
 {
 \bcontraction[2ex]{}{ \mathlarger{\bullet} }{\! \! \!- \! \! \!- \! \!\mathlarger{\bullet}  \! \! \!- \! \! \!- \! \!}{\mathlarger{\bullet}}
 \bcontraction{ \mathlarger{\bullet} \! \! \!- \! \! \!- \! \!}{\mathlarger{\bullet}} { \! \! \!- \! \! \!- \! \!\mathlarger{\bullet}  \! \! \!- \! \! \!- \! \!}{\mathlarger{\bullet}}
\mathlarger{\bullet} \! \! \!- \! \! \!- \! \!\mathlarger{\bullet}  \! \! \!- \! \! \!- \! \!\mathlarger{\bullet} \! \! \!- \! \! \!- \! \!\mathlarger{\bullet}
 }
 %
+  
\sum_{\substack{r_1 \ne r_2 \ne r_3 \\ r_1 \ne r_3}}^n
\overset{\MT_{r_1} \MT_{r_2} \MT_{r_3} \MT_{r_1}}
 {
 \bcontraction{}{ \mathlarger{\bullet} }{\! \! \!- \! \! \!- \! \!\mathlarger{\bullet}  \! \! \!- \! \! \!- \! \!\mathlarger{\bullet} \! \! \!- \! \! \!- \! \!}{\mathlarger{\bullet}}
\mathlarger{\bullet} \! \! \!- \! \! \!- \! \!\mathlarger{\bullet}  \! \! \!- \! \! \!- \! \!\mathlarger{\bullet} \! \! \!- \! \! \!- \! \!\mathlarger{\bullet}
 }
 %
 +
\sum_{r_1 \ne r_2  \ne r_3}^n
\overset{\MT_{r_1} \MT_{r_2} \MT_{r_1} \MT_{r_3}}
 {
 \bcontraction{}{ \mathlarger{\bullet} }{\! \! \!- \! \! \!- \! \!\mathlarger{\bullet}  \! \! \!- \! \! \!- \! \!}{\mathlarger{\bullet}}
 %
\mathlarger{\bullet} \! \! \!- \! \! \!- \! \!\mathlarger{\bullet}  \! \! \!- \! \! \!- \! \!\mathlarger{\bullet} \! \! \!- \! \! \!- \! \!\mathlarger{\bullet}
 }
 + \sum_{r_1 \ne r_2  \ne r_3}^n
\overset{\MT_{r_1} \MT_{r_2} \MT_{r_1} \MT_{r_3}}
 {
 \bcontraction{ \mathlarger{\bullet} \! \! \!- \! \! \!- \! \!}{\mathlarger{\bullet}} { \! \! \!- \! \! \!- \! \!\mathlarger{\bullet}  \! \! \!- \! \! \!- \! \!}{\mathlarger{\bullet}}
\mathlarger{\bullet} \! \! \!- \! \! \!- \! \!\mathlarger{\bullet}  \! \! \!- \! \! \!- \! \!\mathlarger{\bullet} \! \! \!- \! \! \!- \! \!\mathlarger{\bullet}
 }
 %
 + 
\sum_{\substack{r_1 \ne r_2  \ne r_3 \ne r_4 \\ r_3 \ne r_1 \ne r_4 \ne r_2}}^n
 \overset{\MT_{r_1} \MT_{r_2} \MT_{r_3} \MT_{r_4} }
{\mathlarger{\bullet} \! \! \!- \! \! \!- \! \!\mathlarger{\bullet}  \! \! \!- \! \! \!- \! \!\mathlarger{\bullet} \! \! \!- \! \! \!- \! \!\mathlarger{\bullet}}  
\Biggr\} \nonumber \\ + && \cdots.
\label{Transition_Density}
\ee 
Here each circle denotes a scattering event given by the $\widehat{ \mathbb{T} }_r$-matrix of a delta slice, Eq. \eqref{Toperator1S}, the continuous line between two circles, represents the diagonal matrix $-\bm{G}_0(x_{r_1},x_{r_2}) \equiv -\bm{G}_0(1,2)$, whose elements are the channel Green functions $-g_{0}\left(k_{b};x_{r_{1}},x_{r_{2}} \right)=-ie^{ik_{b}\vert x_{r_{1}} - x_{r_{2}} \vert}/2k_{b}$ connecting two different slices,  being $b$ any possible open or closed channel: see text below Eq. \eqref{RelevantDiagrams}.

It is important to notice that Eq. \eqref{Transition_Density}, does represent a series expansion in multiple scattering, where the number of circles in each diagram specifies the order in multiple scattering; however, to obtain the limiting macroscopic statistics of the BB, it is convenient to reorganize the diagrammatic series of Eq. \eqref{Transition_Density}, in terms of restricted sum symbols, each one containing all the multiple scattering contributions between a fix number of scatters; for instance, the second, third and fifth diagrams of Eq. \eqref{Transition_Density} can be grouped into the restricted sum symbol for $r_{1} \neq r_{2}$, which contains all the multiple scattering contributions between two slices. Once the series expansion of $\widehat{ \mathbb{T}  }_{aa_{0}}\left(x^{\prime},x^{\prime\prime}\right)$ is rewritten in this convenient form, 
\begin{subequations}
\be
\MT &=&  
\sum_{r_1}^n \Biggr\{  \overset{\MT_{r_1} }
{\mathlarger{\bullet} } 
\Biggl\}_{[1 \   \text{slice}]} \\
&& + 
\sum_{r_1 \ne r_2}^n \Biggr\{ \overset{\MT_{r_1} \MT_{r_2} }  
{\mathlarger{\bullet} \! \! \!- \! \! \!- \! \!\mathlarger{\bullet} }  
+ 
 \overset{\MT_{r_1} \MT_{r_2} \MT_{r_1}}
 {
 \bcontraction{}{ \mathlarger{\bullet} }{ \! \! \!- \! \! \!- \! \!\mathlarger{\bullet}  \! \! \!- \! \! \!- \! \!}{\mathlarger{\bullet}}
 \mathlarger{\bullet} \! \! \!- \! \! \!- \! \!\mathlarger{\bullet}  \! \! \!- \! \! \!- \! \!\mathlarger{\bullet}
 }
 +  
\overset{\MT_{r_1} \MT_{r_2} \MT_{r_1} \MT_{r_2}}
 {
 \bcontraction[2ex]{}{ \mathlarger{\bullet} }{\! \! \!- \! \! \!- \! \!\mathlarger{\bullet}  \! \! \!- \! \! \!- \! \!}{\mathlarger{\bullet}}
 \bcontraction{ \mathlarger{\bullet} \! \! \!- \! \! \!- \! \!}{\mathlarger{\bullet}} { \! \! \!- \! \! \!- \! \!\mathlarger{\bullet}  \! \! \!- \! \! \!- \! \!}{\mathlarger{\bullet}}
\mathlarger{\bullet} \! \! \!- \! \! \!- \! \!\mathlarger{\bullet}  \! \! \!- \! \! \!- \! \!\mathlarger{\bullet} \! \! \!- \! \! \!- \! \!\mathlarger{\bullet}
 }
 +
 +
\overset{\MT_{r_1} \MT_{r_2} \MT_{r_1} \MT_{r_2} \MT_{r_1} }
 {
 \bcontraction{ \mathlarger{\bullet} \! \! \!- \! \! \!- \! \!}{\mathlarger{\bullet}} { \! \! \!- \! \! \!- \! \!\mathlarger{\bullet}  \! \! \!- \! \! \!- \! \!}{\mathlarger{\bullet}}
 \bcontraction[2ex]{}{ \mathlarger{\bullet} }{\! \! \!- \! \! \!- \! \!\mathlarger{\bullet}  \! \! \!- \! \! \!- \! \!}{\mathlarger{\bullet}}
 \bcontraction[2ex]{\mathlarger{\bullet} \! \! \!- \! \! \!- \! \!\mathlarger{\bullet} \! \! \!- \! \! \!- \! \!
 }
 { \mathlarger{\bullet} }{\! \! \!- \! \! \!- \! \!\mathlarger{\bullet}  \! \! \!- \! \! \!- \! \!}{\mathlarger{\bullet}}
\mathlarger{\bullet} \! \! \!- \! \! \!- \! \!\mathlarger{\bullet}  \! \! \!- \! \! \!- \! \!\mathlarger{\bullet} \! \! \!- \! \! \!- \! \!\mathlarger{\bullet}  \! \! \!- \! \! \!- \! \!\mathlarger{\bullet}
 }
 + \cdots \Biggl\}_{[2 \   \text{slices}]} \\
 && + 
 %
%
\sum_{\substack{r_1 \ne r_2 \ne r_3 \\ r_1 \ne r_3}}^n \Biggr\{
 \overset{\MT_{r_1} \MT_{r_2} \MT_{r_3}}
{\mathlarger{\bullet} \! \! \!- \! \! \!- \! \!\mathlarger{\bullet}  \! \! \!- \! \! \!- \! \!\mathlarger{\bullet}}  
+   
\overset{\MT_{r_1} \MT_{r_2} \MT_{r_3} \MT_{r_1}}
 {
 \bcontraction{}{ \mathlarger{\bullet} }{\! \! \!- \! \! \!- \! \!\mathlarger{\bullet}  \! \! \!- \! \! \!- \! \!\mathlarger{\bullet} \! \! \!- \! \! \!- \! \!}{\mathlarger{\bullet}}
\mathlarger{\bullet} \! \! \!- \! \! \!- \! \!\mathlarger{\bullet}  \! \! \!- \! \! \!- \! \!\mathlarger{\bullet} \! \! \!- \! \! \!- \! \!\mathlarger{\bullet}
 }
  +
\overset{\MT_{r_1} \MT_{r_2} \MT_{r_1} \MT_{r_3}}
 {
 \bcontraction{}{ \mathlarger{\bullet} }{\! \! \!- \! \! \!- \! \!\mathlarger{\bullet}  \! \! \!- \! \! \!- \! \!}{\mathlarger{\bullet}}
 %
\mathlarger{\bullet} \! \! \!- \! \! \!- \! \!\mathlarger{\bullet}  \! \! \!- \! \! \!- \! \!\mathlarger{\bullet} \! \! \!- \! \! \!- \! \!\mathlarger{\bullet}
 }
 +
\overset{\MT_{r_1} \MT_{r_2} \MT_{r_1} \MT_{r_3}}
 {
 \bcontraction{ \mathlarger{\bullet} \! \! \!- \! \! \!- \! \!}{\mathlarger{\bullet}} { \! \! \!- \! \! \!- \! \!\mathlarger{\bullet}  \! \! \!- \! \! \!- \! \!}{\mathlarger{\bullet}}
\mathlarger{\bullet} \! \! \!- \! \! \!- \! \!\mathlarger{\bullet}  \! \! \!- \! \! \!- \! \!\mathlarger{\bullet} \! \! \!- \! \! \!- \! \!\mathlarger{\bullet}
 } 
 + \cdots
 \Biggl\}_{[3 \   \text{slices}]} \\
 && 
 + 
 \sum_{\substack{r_1 \ne r_2  \ne r_3 \ne r_4 \\ r_3 \ne r_1 \ne r_4 \ne r_2}}^n
 \Biggr\{ \overset{\MT_{r_1} \MT_{r_2} \MT_{r_3} \MT_{r_4} }
{\mathlarger{\bullet} \! \! \!- \! \! \!- \! \!\mathlarger{\bullet}  \! \! \!- \! \! \!- \! \!\mathlarger{\bullet} \! \! \!- \! \! \!- \! \!\mathlarger{\bullet}}  
+ \cdots \Biggr\}_{[4 \   \text{slices}]}  + \cdots
\label{Transition_Density_rewritten}
\ee
\end{subequations}
the resulting expression is introduced in Eq. \eqref{ReDef_Back_For_scatt_Ampli} to calculate the transition matrix elements 
\begin{eqnarray}
\mathtt{T} _{aa_{0}} ^ {\sigma^{\prime} \sigma} &=& 
 \sum_{r_1}^n  \varphi_{\sigma^{\prime}} ^{\ast} \left(k_{a};x_{r_1}\right) 
 \Biggr\{
 \bm{\mathcal{T}}_{r_1}
\Biggl\}_{a a_0}
\varphi_{\sigma} \left(k_{a_{0}};x_{r_1}\right) \nonumber \\
& +&  
 \sum_{r_1 \ne r_2}^n  \varphi_{\sigma^{\prime}} ^{\ast} \left(k_{a};x_{r_1}\right) 
 \Biggr\{-
 \bm{\mathcal{T}}_{r_1} \bm{G}_0(1,2)  \bm{\mathcal{T}}_{r_2} -
 \bm{\mathcal{T}}_{r_1} \bm{G}_0(1,2)  \bm{\mathcal{T}}_{r_2}  \bm{G}_0(2,1)  \bm{\mathcal{T}}_{r_1}
 \bm{G}_0(1,2)  \bm{\mathcal{T}}_{r_2}  + \cdots
\Biggl\}_{a a_0}
\varphi_{\sigma} \left(k_{a_{0}};x_{r_2}\right) \nonumber \\
& +&  
 \sum_{r_1 \ne r_2}^n  \varphi_{\sigma^{\prime}} ^{\ast} \left(k_{a};x_{r_1}\right) 
 \Biggr\{
 \bm{\mathcal{T}}_{r_1} \bm{G}_0(1,2)  \bm{\mathcal{T}}_{r_2}  \bm{G}_0(2,1)  \bm{\mathcal{T}}_{r_1}
 + \cdots
\Biggl\}_{a a_0}
\varphi_{\sigma} \left(k_{a_{0}};x_{r_1}\right) +
\sum_{\substack{r_1 \ne r_2 \ne r_3 \\ r_1 \ne r_3}}^n  \cdots
\\
-\frac{i}{2} && \! \! \! \mathtt{T} _{aa_{0}} ^ {\sigma^{\prime} \sigma} =
 \sum_{r_1}^n  e^{-i\sigma'k_a x_{r_1}} 
 \Biggr\{
 (\tilde{\bm{r}}_{r_1})_{a a_0}
\Biggl\}
e^{i\sigma k_{a_0} x_{r_1}}  \nonumber \\
& +&  
 \sum_{r_1 \ne r_2}^n   e^{-i\sigma'k_a x_{r_1}}  
\Biggr\{  \sum_{b_1=1}^\infty \left(
(\tilde{\bm{r}}_{r_1})_{a b_1} e^{ik_{b_1}|x_{r_1}-x_{r_2}|} (\tilde{\bm{r}}_{r_2})_{b_1 a_0}\right)  + \cdots
\Biggl\}
e^{i\sigma k_{a_0} x_{r_2}}   \nonumber \\
& +&  
\sum_{r_1 \ne r_2}^n   e^{-i\sigma'k_a x_{r_1}} 
 \Biggr\{ \sum_{b_1,b_2}^\infty \left(
 (\tilde{\bm{r}}_{r_1})_{a b_1} e^{ik_{b_1}|x_{r_1}-x_{r_2}|} (\tilde{\bm{r}}_{r_2})_{b_1 b_2} 
 e^{ik_{b_2}|x_{r_2}-x_{r_1}|} (\tilde{\bm{r}}_{r_1})_{b_2 a_0} \right)
 + \cdots
\Biggl\}
e^{i\sigma k_{a_0} x_{r_1}}  + \sum_{\substack{r_1 \ne r_2 \ne r_3 \\ r_1 \ne r_3}}^n  \cdots \nonumber \\
\label{TMwithR}
\end{eqnarray}
where we have made use of the definition of the generalised reflection matrix $\tilde{\bm{r}}_r$, Eq. \eqref {GenRefMat}. 
The statistical average of the series expansion, Eq. \eqref{TMwithR}, leads to the  diagrammatic expansion Eq. \eqref{RelevantDiagrams}.

\section{\label{Avtra} Averaged transmission amplitudes}

In order to understand the effects of the SWLA and the inclusion or omission of the evanescent modes in the macroscopic statistics, we use the perturbative technique of Sec. \ref{Moments_of_TransitionMatrix}, to obtain, the average of the diagonal transmission matrix amplitudes of a disordered wire. From Eq. \eqref{RelevantDiagrams2},
and in the DWSL, Eq. \eqref{Weak_dense_limits}, we have 
\begin{eqnarray}
\left\langle t_{a_{0}a_{0}}\right\rangle _{L} &=& 1-\frac{i}{2} \left\langle \mathtt{T}^{+ +}_{a_{0}a_{0}}\right\rangle _{L} 
\nonumber \\
&=& 
1+ f\alpha_{ a_0 a_0}^{(1)}
\left[\int_0^L d x_{1}  \right]_1 
 +   \sum_{b_1=1}^\infty f\alpha_{ a_0 b_1}^{(1)} f\alpha_{ b_1 a_0}^{(1)}
 \left[\iint_0^L dx_{1} dx_{2} e^{-ik_{a_0} x_{1}} e^{ik_{b_1}|x_{1}-x_{2}|} e^{ik_{a_0} x_{2}}\right]_2 + \cdots 
 \label{Series_Diagonal_Aver_ta0a0} \\
%
&=&
1+ \biggl[
f \alpha_{a_0 a_{0}}^{\left(1\right)}  L\biggl]_1
+ \biggl[ \frac{1}{2!} \left( f\alpha_{a_0a_{0}}^{\left(1\right)} L\right)^2
+ \frac{1}{2} \left(\frac{f \alpha_{a_{0}a_0}^{\left(1\right)}}{ 2ik_{a_{0}} }\right)^2
\left(
e^{2ik_{a_{0}}L} -1-2ik_{a_{0}}L
\right)
\biggr]_2^{(b_1 = a_0)}
\nonumber
\\
&+&
\underset{{ b_{1}\neq a_{0} }}{\sum_{b_{1}=1}^{\infty}}
f \alpha_{a_{0}b_{1}}^{\left( 1\right) }f \alpha_{b_{1}a_{0}}^{\left( 1\right) }
\left[
\frac{e^{i\left(k_{b_{1}}-k_{a_{0}} \right)L} -1-i\left(k_{b_{1}}-k_{a_{0}} \right)L}
{ \left[ i\left( k_{b_{1}} - k_{a_{0}}\right)\right]^{2}}
+
\frac{e^{i\left(k_{b_{1}}+k_{a_{0}} \right)L} -1-i\left(k_{b_{1}}+k_{a_{0}} \right)L}{ \left[ i\left(k_{b_{1}}+k_{a_{0}} \right)\right]^{2}}
\right]_2
+ \cdots . \label{Series_Diagonal_Aver_ta0a02}
\end{eqnarray}
\end{widetext}
In the second row of Eq. \eqref{Series_Diagonal_Aver_ta0a0}, we have written the single scattering contributions $[\cdots]_1$ and the first term of the multiple scattering contributions corresponding to two scatterers, $[\cdots]_2$. The first row of Eq.
 \eqref{Series_Diagonal_Aver_ta0a02} shows the terms where, up to second order in multiple scattering, only the incoming open channel $a_{0}$ appears in the direct channel-channel transitions $a_{0}\rightarrow a_{0}$: the term proportional $\alpha_{a_{0}a_0}^{\left(1\right)}$, is the first order in multiple scattering, while those terms proportional to $(\alpha_{a_{0}a_0} ^{\left( 1\right)})^{2}$ are second order scattering process, in which the first and the second scattering events are channel-channel transitions $a_{0} \rightarrow a_{0}$. 
If we now consider the SWLA, Eq. \eqref{MyPHDTEq4.43}, the contributions
\begin{equation}
\left(\frac{ \alpha_{a_{0}}^{\left(1\right)}}{ 2ik_{a_{0}} }\right)
\alpha_{a_{0}}^{\left(1\right)} L
, \;\;\;\;
\left(\frac{ \alpha_{a_{0}}^{\left(1\right)}}{ 2ik_{a_{0}} }\right)^2
\left(
e^{2ik_{a_{0}}L} -1
\right)
\label{2ndRowNegliTermsP}
\end{equation}
\noindent are of the order $\sim 1/k\ell$ and $\sim 1/(k\ell)^{2}$, respectively; therefore, in the SWLA $k\ell \gg 1$, the contributions of Eq. \eqref{2ndRowNegliTermsP}, are negligible in comparison to the terms \begin{small}$\alpha_{a_{0}}^{\left(1\right)}  L$\end{small}, \begin{small}$(\alpha_{a_{0}}^{\left(1\right)} L )^2/2!$\end{small}, which are of the order $\sim 1/(k\ell)^{0}$. 
 
The second row of Eq. \eqref{Series_Diagonal_Aver_ta0a02}, contains all those double scattering terms, $\sim \alpha_{a_{0}b_{1}}^{\left( 1\right) } \alpha_{b_{1}a_{0}}^{\left( 1\right) }$, where the channel-channel transition $a_{0} \rightarrow a_{0}$ takes place through any intermediate open channel $1 \leq b_{1} \leq N$ ($b_{1} \neq a_{0}$), i.e., through propagating transport, but also through closed channels $b_{1} >N$, i.e., through evanescent transport. Obviously, the contribution of evanescent modes is exponentially small $e^{ik_{b_{1}}L}\rightarrow e^{-\kappa_{b_{1}}L}$ and can be neglected. The remaining propagating transport contributions of the second row of  Eq. \eqref{Series_Diagonal_Aver_ta0a02}, are again of the order $\sim 1/k\ell$ and $\sim 1/(k\ell)^{2}$, so they can also be neglected in the SWLA. An equivalent analysis shows that non-diagonal terms  $\left\langle t_{a a_{0}}\right\rangle _{L}$ are of the order of $\sim 1/(k\ell)$. In this simple case, it is possible to sum the whole series of {\em the dominant contributions} [which are of the order $\sim 1/(k\ell)^{0}$], giving rise to the Eq. \eqref{AverTransAmplitude}.
\bibliographystyle{apsrev4-1}
\bibliography{YepezSaenzBibtex}

\end{document}